\documentclass[12pt]{iopart}
\eqnobysec
\usepackage[mathscr]{euscript}
\expandafter\let\csname equation*\endcsname\relax
\expandafter\let\csname endequation*\endcsname\relax
\usepackage{amsmath, amsthm, amssymb}
\usepackage{enumerate}
\usepackage{setspace}
\usepackage{natbib}

\newtheorem{theorem}{Theorem}[section]

\newtheorem{proposition}[theorem]{Proposition}

\newcommand{\derv}[1]{\frac{\partial}{\partial #1}}

\newcommand{\deriv}[2]{\frac{\partial #1}{\partial #2}}

\newcommand{\beqn}{\begin{equation}}
\newcommand{\eeqn}{\end{equation}}
\newcommand{\beqnar}{\begin{eqnarray}}
\newcommand{\eeqnar}{\end{eqnarray}}

\usepackage{hyperref}
\begin{document}

\title[On MHD Field Theory]
{On Magnetohydrodynamic Gauge Field Theory} 
\author{G.M. Webb${}^1$ and S.C. Anco${}^2$} 
\address{${}^1$ Center for Space Plasma and Aeronomic Research, The University of Alabama in Huntsville, 
 Huntsville AL 35805, USA}

\address{${}^2$Department of Mathematics, Brock University, 
St. Catharines, ON L2S 3A1 Canada}

 
\ead{gmw0002@uah.edu}


\begin{abstract}
Clebsch potential gauge field theory for magnetohydrodynamics is developed based 
in part on the theory of \cite{Calkin63}. It is shown how the polarization vector 
${\bf P}$ in Calkin's approach naturally arises from the Lagrange multiplier 
constraint equation for Faraday's equation for the magnetic induction ${\bf B}$, 
or alternatively from the magnetic vector potential form of Faraday's equation.
Gauss's equation, (divergence of ${\bf B}$ is zero) is 
incorporated in the variational principle
by means of a Lagrange multiplier constraint.
Noether's theorem coupled with the gauge symmetries is used to derive the conservation 
laws for (a)\ magnetic helicity, (b)\ cross helicity, (c) fluid helicity for 
non-magnetized fluids, and (d) a class of conservation laws associated with curl 
and divergence equations which applies to Faraday's equation and Gauss's equation.
The magnetic helicity conservation law is due to a gauge symmetry in MHD and not 
due to a fluid relabelling symmetry. 
The analysis is carried out for the general case of a non-barotropic gas in which 
the gas pressure and internal energy density depend on both the entropy $S$
and the gas density $\rho$. The cross helicity and fluid helicity conservation 
laws in the non-barotropic case are nonlocal conservation laws that reduce to 
local conservation laws for the case of a barotropic gas. The connections between 
gauge symmetries, Clebsch potentials  and Casimirs are developed. It is shown that
the gauge symmetry functionals in the work of \cite{Henyey82} satisfy the Casimir 
determining equations.

\end{abstract}

\submitto{{\it J. Phys. A., Math. and Theor.},\today}

\maketitle


\section{Introduction}
In this paper we investigate the role of gauge symmetries and gauge transformations in 
magnetohydrodynamics (MHD) 
based in part on the work of \cite{Calkin63}. \cite{Tanehashi15} 
used the MHD Casimirs of magnetic helicity 
and cross helicity for MHD and the fluid helicity to determine 
 gauge transformations. They used  Clebsch representations 
for the fluid velocity and the magnetic field and related    
 the gauge transformations 
to the non-canonical Poisson bracket for MHD of \cite{MorrisonGreene80, MorrisonGreene82}. 
Our aim is to provide a description of gauge symmetries in MHD using a constrained variational
principle, in which the mass, entropy, Lin constraint, and Faraday's law are included 
in the variational principle using Lagrange multipliers. We include a Lagrange multiplier
term to ensure $\nabla{\bf\cdot}{\bf B}=0$  where ${\bf B}$ is the magnetic field induction.  

The MHD model of \cite{Calkin63} is different from most MHD formulations, since it 
takes into account the polarization ${\bf P}$ and polarization charge density 
$\rho_c$  of the plasma, as well as quadratic electric field terms in the Lagrangian 
(i.e. it includes displacement current effects). In particular, \cite{Calkin63} 
uses the charge continuity equation:
\begin{equation}
\deriv{\rho_c}{t}+\nabla{\bf\cdot}\left({\bf J}+\rho_c {\bf u}\right)=0, \label{eq:gauge1.1}
\end{equation}
where  ${\bf u}$ is the fluid velocity and  $\rho_c$ is the charge density. The charge density $\rho_c$ 
and non-advected current ${\bf J}$ are related to 
the polarization ${\bf P}$ through the equations:
\begin{align}
{\bf J}=&\deriv{\bf P}{t}-\nabla\times({\bf u}\times{\bf P})+{\bf u}\nabla{\bf\cdot}{\bf P}, 
\label{eq:gauge1.2}\\
\nabla{\bf\cdot}{\bf P}=&-\rho_c. \label{eq:gauge1.3}
\end{align}
Note that (\ref{eq:gauge1.1})-(\ref{eq:gauge1.3}) are consistent. Taking the divergence
of (\ref{eq:gauge1.2}) and taking into account (\ref{eq:gauge1.3}) gives the charge 
continuity equation (\ref{eq:gauge1.1}). 
A derivation of (\ref{eq:gauge1.2}) is given in \cite{Panofsky64}. In the  
non-relativistic MHD model the quasi-neutrality of the plasma 
(the plasma approximation) is invoked, and $\rho_c$ is set equal to zero. Calkin retains 
the electric field energy density in his Lagrangian, which is related to the 
displacement current, which is usually neglected in non-relativistic MHD.  
The result (\ref{eq:gauge1.2}) may be written as:
\begin{equation}
\left(\derv{t}+{\mathcal L}_{\bf u}\right) \left( {\bf P}{\bf\cdot} d{\bf S}\right)\equiv 
\frac{d}{dt}\left( {\bf P}{\bf\cdot} d{\bf S}\right)={\bf J}{\bf\cdot} d{\bf S}, 
\label{eq:gauge1.4}
\end{equation}
where ${\cal L}_{\bf u}={\bf u}{\bf\cdot}\nabla$ is the Lie derivative following the flow,  
and $d/dt=\partial/\partial t+{\bf u} {\bf\cdot}\nabla$ is the Lagrangian time 
derivative following the flow.
The quantity:
\begin{equation}
{\bf P}{\bf\cdot}d{\bf S}=P_x dy\wedge dz+P_y dz\wedge dx+P_z dx\wedge dy, \label{eq:gauge1.5}
\end{equation} 
is the polarization 2-form. In the usual MHD, non-relativistic limit $\nabla{\bf\cdot}{\bf P}=0$ 
(i..e. $\rho_c=0$), and (\ref{eq:gauge1.2}) simplifies to:
\begin{equation}
{\bf J}=\deriv{\bf P}{t}-\nabla\times({\bf u}\times{\bf P}), \label{eq:gauge1.6}
\end{equation}
which resembles Faraday's equation (but for ${\bf P}$ rather than ${\bf B}$), in which 
${\bf J}\neq 0$ is the source term. We show that (\ref{eq:gauge1.2}) or (\ref{eq:gauge1.6})
 arises as the evolution equation for a vector Lagrange multiplier 
used in enforcing Faraday's equation in the variational principle.

 Magnetic helicity is a key quantity in describing the topology 
of magnetic fields (e.g. \cite{Moffatt69, Moffatt78}, \cite{Moffatt92}, \cite{Berger84}, \cite{Finn85}, 
\cite{Arnold98}). 
The magnetic helicity $H_M$ is defined as:
\begin{equation}
H_M=\int_V\ \boldsymbol{\omega}_1\wedge d\boldsymbol{\omega}_1=\int_Vd^3x\ {\bf A}{\bf\cdot}{\bf B}, 
\label{eq:gauge1.7}
\end{equation}
where $\boldsymbol{\omega}_1={\bf A}{\bf\cdot}d{\bf x}$ is the magnetic vector potential 1-form, 
$d\boldsymbol{\omega}_1={\bf B}{\bf\cdot}d{\bf S}$ is the magnetic field 2-form; ${\bf B}=\nabla\times{\bf A}$ is the magnetic induction, ${\bf A}$ is the magnetic vector potential
and $V$ is the isolated volume containing the field. Magnetic helicity is an invariant of MHD 
(e.g. \cite{Elsasser56},
\cite{Woltjer58}, \cite{Moffatt69, Moffatt78}). In (\ref{eq:gauge1.7}) it is assumed 
that the normal magnetic field component $B_n={\bf B}{\bf\cdot}{\bf n}$ vanishes on the boundary $\partial V$. 
The first form of magnetic helicity in (\ref{eq:gauge1.7}) is known as the Hopf invariant. 
It is also a Casimir of MHD.  A Casimir C is a functional that has the property $\{F,C\}=0$, 
for a general functional $F$, where $\{F,G\}$ is the Hamiltonian Poisson bracket for the Hamiltonian system (e.g. 
\cite{Padhye96a, Padhye96b}, \cite{Morrison98}, \cite{Hameiri04}).

Cross helicity in MHD is defined as the integral:
\begin{equation}
H_C=\int_V d^3x\ {\bf u}{\bf\cdot}{\bf B}. \label{eq:gauge1.8}
\end{equation}
In (\ref{eq:gauge1.8}) it is assumed that
${\bf B}{\bf\cdot}{\bf n}=0$ on the boundary $\partial V$ of the volume $V$.
It is a Casimir for barotropic MHD (e.g. \cite{Padhye96a,Padhye96b}). It is also referred to as a rugged invariant 
in MHD turbulence theory (e.g. \cite{Matthaeus82}). For a barotropic equation of state for the gas 
$dH_C/dt=0$ for a volume $V=V_m$ moving with the flow. For a non-barotropic equation of state for the gas, 
\begin{equation}
\frac{dH_C}{dt}=\int_V d^3x\ T{\bf B}{\bf\cdot}\nabla S, \label{eq:gauge1.9}
\end{equation}
which reduces to $dH_C/dt=0$  for a barotropic gas. 
One can define a modified form of cross helicity for a non-barotropic gas as:
\begin{equation}
H_{CNB}=\int_V d^3x\ ({\bf u}+r\nabla S){\bf\cdot} {\bf B},\quad \hbox{where}\quad \frac{dr}{dt}=-T({\bf x},t), 
 \label{eq:gauge1.10}
\end{equation}
where $T$ is the temperature of the gas (e.g. \cite{Webb14a, Webb14b}, \cite{Yahalom16, Yahalom17a, Yahalom17b}). 
Note that 
\begin{equation}
\frac{dH_M}{dt}=0,\quad \frac{dH_{CNB}}{dt}=0. \label{eq:gauge1.11}
\end{equation}
 
 \cite{Webb14a, Webb14b} derived MHD conservation laws (CL's) using the Lie dragging approach of 
\cite{Tur93} and also by using Noether's theorems. In particular, they derived the cross 
helicity CL for both barotropic and non-barotropic MHD. 
\cite{Yahalom16, Yahalom17a, Yahalom17b}
\cite{Yahalom13} investigated the magnetic helicity and cross helicity conservation laws
for both barotropic and non-barotropic equations of state for the gas. He gave  
  interpretations of these conservation 
laws in terms of  generalized Aharonov-Bohm effects.     
The Aharonov-Bohm effect describes the change in phase of the wave function 
in quantum mechanics, depending on the path integral of $\int {\bf A}{\bf\cdot}d{\bf x}$ about an isolated 
magnetic flux in the vicinity of the path (\cite{Aharonov59}).
 Yahalom expresses his results in terms of the
magnetic helicity per unit magnetic flux, and the cross helicity per unit magnetic flux
and in terms of the magnetic metage ( metage is a measure of distance along a field line 
which is the intersection of two Euler surfaces, see e.g. Appendix E).  

These conservation laws provides tests for symplectic MHD codes (see \cite{Krauss16} for reduced MHD, 
and \cite{Krauss17} for 2D MHD),  which are designed to preserve magnetic helicity, cross helicity and energy,
and are sometimes known as geometric integrators. They also preserve the Hamiltonian Poisson bracket 
structure of the equations.  
A brief discussion of Yahalom's 
 approach is described 
in Appendix E. 
 \cite{Yahalom13,Yahalom17a,Yahalom17b} provides a semi-analytical example where 
the magnetic helicity per unit magnetic flux is related to the jump of the metage potential 
as one moves along a field line formed by the intersection of two Euler potential surfaces. 
This type of example is in principle very useful for the testing of MHD codes, against 
theoretical ideas of magnetic  helicity and Clebsch potentials. 
The conservation law for non-barotropic cross helicity, provides a simple challenge for MHD
codes, since it can be formulated in the context of the usual MHD equations, by the addition
of a single non-local variable $r$ which satisfies $dr/dt\equiv r_t+{\bf u}{\bf\cdot} \nabla r=-T({\bf x},t)$
where $T({\bf x},t)$ is the temperature of the gas (Yahalom (2017a,b) uses $\sigma=-r$). There is also a 
fluid dynamical analog of this conservation law in ideal fluid dynamics in which ${\bf B}$ is replaced 
by the generalized vorticity $\boldsymbol{\Omega}=\boldsymbol{\omega}+\nabla r\times\nabla S$ where 
$\boldsymbol{\omega}=\nabla\times{\bf u}$ is the fluid vorticity (e.g. \cite{Mobbs81}). 
Testing the magnetic helicity conservation 
law requires the calculation of the magnetic vector potential ${\bf A}$, which depends on the gauge. 
To obtain a physically  meaningful topological result requires further analysis (e.g. \cite{Berger84}, 
\cite{Moffatt92}, \cite{Arnold98}, \cite{Webb10a}). 

Section 2 introduces the Eulerian variational principle, using Lagrange multipliers
to enforce the mass, entropy advection, Lin constraints, Gauss's law,
 and Faraday's law. 
 Section 3, gives the determining equations for gauge symmetries.  
The main idea is that the gauge symmetries do not alter the physical variables. 
 A central equation is the Clebsch potential 
expansion for the fluid velocity ${\bf u}$, which follows from requiring that the action is stationary 
for Eulerian variations $\delta{\bf u}$ of the fluid velocity. 

Section 3 investigates the conditions for Lie transformations to leave the 
action and the physical variables invariant  up to a divergence transformation
by using the approach of \cite{Calkin63}. 

Section 4 describes Noether's theorem, which requires the action remain 
invariant   up to a divergence transformation. 
This condition gives conservation laws of the MHD system due to  
the gauge symmetries. We derive the magnetic helicity conservation law, 
and the cross helicity conservation laws using Noether's theorem. 
These two conservation 
laws were derived by \cite{Calkin63}. 
Our formulation is slightly 
more general than \cite{Calkin63}, who considered 
 the case of a barotropic gas for which the pressure $p=p(\rho)$ whereas we use 
 a non-barotropic gas for which $p=p(\rho,S)$ where $\rho$ is the density 
and $S$ is the entropy of the gas.  For a barotropic gas the cross helicity conservation 
law is a local conservation, but  
 it is a nonlocal conservation law for a non-barotropic gas. 

Section 5 explores the connection between gauge symmetries and Casimirs based on 
the gauge field theory of \cite{Henyey82}. We show that the  condition 
that the fluid velocity remain invariant coupled with the 
conditions that the density, entropy and magnetic field and Lin constraint variables 
$\mu^k$ do not change under gauge transformations  implies that 
the gauge transformation generating functionals are Casimirs.  

Section 6 concludes with a summary and discussion. 

Some of the important results established in the paper are: (\romannumeral1)\ 
A version of Noether's theorem describing gauge symmetries in MHD; (\romannumeral2)\ 
A gauge symmetry associated with the Lagrange multiplier $\boldsymbol{\Gamma}$ that 
enforces Faraday's equation, and the Lagrange multiplier 
$\nu$ that enforces $\nabla{\bf\cdot}{\bf B}=0$ give rise via Noether's theorem 
to the magnetic helicity conservation law (this symmetry is not a fluid relabelling symmetry). 
\cite{Calkin63} derived the magnetic helicity conservation law 
for a more complicated version 
of MHD involving the polarization vector ${\bf P}$ (for usual MHD $\nabla{\bf\cdot}{\bf P}=0$); 
(\romannumeral3)\ An analysis of the gauge symmetry determining equations (Appendix B) 
shows the connection of these equations to the Lie symmetry structure of 
the steady MHD equations established by \cite{Bogoyavlenskij02}, and \cite{Schief03}; 
(\romannumeral4)\ the gauge symmetry functionals of \cite{Henyey82} 
are shown to satisfy the Casimir determining equations (Appendix D);
(\romannumeral5)\ A gauge symmetry associated with $\boldsymbol{\Gamma}$ and $\nu$ 
and an arbitrary potential $\Lambda({\bf x},t)$ is related to fluid conservation 
laws obtained by \cite{Cheviakov14}. Cheviakov studied conservation laws for fluid models 
containing an equation subsystem of the form ${\bf N}_t+\nabla\times{\bf M}=0$  
and $\nabla{\bf\cdot M}=0$. A nonlocal conservation law with conserved density 
$D={\bf B}{\bf\cdot}\nabla\phi$ ($\phi$ is the potential in Bernoulli's equation) 
is derived. This method can be used to obtain the potential vorticity 
conservation law for MHD of \cite{WebbMace15}). (\romannumeral6)\ The  
 mass continuity equation, 
the Eulerian entropy conservation equation, Gauss's equation and Faraday's equation 
may be obtained by using Noether's theorem and gauge symmetries. The mass continuity equation 
can be regarded as due to symmetry breaking of a fluid relabelling symmetry (\cite{Holm98}).

Detailed calculations are given in  Appendices A-E. Appendix A derives the MHD momentum
equation from the Clebsch variational equations. Appendices B and C derive solutions
of the gauge symmetry equations. In particular, Appendix B gives solutions of
the equations analogous to the steady state Faraday equation, Gauss law
and the  mass continuity equation. Faraday's equation may be described
by a 2 dimensional simple Abelian Lie algebra which is integrable by Frobenius theorem
(cf. \cite{Bogoyavlenskij02},  \cite{Schief03}, \cite{Webb05b}). An alternative solution
method was presented in Appendix C. Appendix D shows that the variational equations
for gauge potential functionals $F$ (e..g \cite{Henyey82})  are equivalent to the
Casimir determining equations (e.g. \cite{Hameiri04}). Appendix E describes the work of 
\cite{Yahalom16, Yahalom17a, Yahalom17b}. 

\section{MHD variational principles}
There are a variety of different variational principles that describe MHD. 
We use the action:
\begin{equation}
{\cal A}=\int \ell\ d^3x\ dt, \label{eq:gauge2.1}
\end{equation}
where
\begin{align}
\ell=&\left(\frac{1}{2}\rho u^2-\varepsilon(\rho,S)-\frac{B^2}{2\mu}-\rho\Phi({\bf x})\right) 
+\phi\left(\deriv{\rho}{t}+\nabla{\bf\cdot}(\rho {\bf u})\right)\nonumber\\
&+\beta\left(\deriv{S}{t}+{\bf u}{\bf\cdot}\nabla S\right)+ \sum_k \lambda^k 
\left(\deriv{\mu^k}{t}+{\bf u}{\bf\cdot}\nabla \mu^k\right)\nonumber\\
&+\nu\nabla{\bf\cdot}{\bf B}
+\boldsymbol{\Gamma}{\bf\cdot}\left(\deriv{\bf B}{t}
-\nabla\times\left({\bf u}\times {\bf B}\right)+\epsilon_1 {\bf u}\nabla{\bf\cdot}{\bf B}\right). 
\label{eq:gauge2.2}
\end{align}
The Lagrangian (\ref{eq:gauge2.2}) consists of the kinetic minus the potential energy of the
MHD fluid, including the gravitational potential energy term $\rho\Phi({\bf x})$
due to an external gravitational field (e.g. in stellar wind theory, $\Phi({\bf x})$ is 
the gravitational potential energy due to the star). The other terms in (\ref{eq:gauge2.2}) 
are constraint terms where $\phi$, $\beta$, $\lambda^k$ ($k<3$), $\nu$ 
and $\boldsymbol{\Gamma}$ are  Clebsch potentials that enforce mass, entropy advection, 
Lin constraint, Gauss's law ($\nabla{\bf\cdot}{\bf B}=0$) and Faraday's law in the MHD limit.

 Lin constraints (\cite{Lin63}, \cite{Cendra87}) were originally introduced in fluid dynamics
to describe fluid vorticity in Clebsch expansions for the velocity ${\bf u}$ 
of the form:
\begin{equation} 
{\bf u}=\nabla\phi-r\nabla S-\sum_{k=1}^3 \lambda^k\nabla\mu^k, \label{eq:gauge2.2a}
\end{equation}
where $r=\beta/\rho$. 
The necessity of the terms $-\sum_{k=1}^3 \lambda^k\nabla\mu^k$ in the expansion becomes obvious
when one calculates the vorticity of the fluid, by taking the curl of ${\bf u}$, to obtain
\begin{equation}
\boldsymbol{\omega}=\nabla\times{\bf u}=-\nabla r\times \nabla S-\sum_{k=1}^3\nabla \lambda^k\times\nabla\mu^k.
\label{eq:gauge2.2b}
\end{equation}
From (\ref{eq:gauge2.2b}) the term $-\nabla r\times\nabla S$  represents the vorticity induced 
by the non-barotropic effects in which the fluid density and pressure gradients are misaligned (alternatively
fluid spin is produced by the misalignment of the temperature an entropy gradients (e.g. \cite{Pedlosky87})). 
However, the fluid element could also have non-zero vorticity due to the initial conditions.
This latter source of vorticity is represented by the sum over $k$ from $k=1$ to $k=3$ in (\ref{eq:gauge2.2b}).
Further analysis (e.g. \cite{Cendra87}, \cite{Holm98}) links the Clebsch expansion for ${\bf u}$ 
with a fluid relabeling diffeomorphism and a momentum map in which the Lagrange multipliers and the 
constraint equation variables are canonically conjugate variables (see also \cite{Zakharov97}).

We use the variational principle (\ref{eq:gauge2.1})-(\ref{eq:gauge2.2}) 
both with $\epsilon_1=0$ in Sections 3 and 4, and with $\epsilon_1=1$ in Section 5. 
The case $\epsilon_1=0$ gives the usual form of Faraday's equation for the case 
$\nabla{\bf\cdot}{\bf B}=0$. The case $\epsilon_1=1$ and $\nu=0$ 
is useful in exploring the effect of 
$\nabla{\bf\cdot}{\bf B}\neq 0$, which is of interest in numerical MHD codes, 
in which $\nabla{\bf\cdot}{\bf B}\neq 0$ may be generated by the approximation scheme 
(e.g. \cite{Powell99}, \cite{Janhunen00}, \cite{Dedner02}, \cite{Webb10b}).

In (\ref{eq:gauge2.2}) Faraday's equation is written in the form:
\begin{equation}
\deriv{\bf B}{t}-\nabla\times\left({\bf u}\times {\bf B}\right)
+\epsilon_1 {\bf u}\nabla{\bf\cdot}{\bf B}=0, \label{eq:gauge2.2aa}
\end{equation}
For the case $\epsilon_1=1$, equation (\ref{eq:gauge2.2aa}) is equivalent to the equation:
\begin{equation}
\left(\derv{t}+{\cal L}_{\bf u}\right) {\bf B}{\bf\cdot} d{\bf S}=
\left(\deriv{\bf B}{t}-\nabla\times\left({\bf u}\times {\bf B}\right)
+{\bf u}\nabla{\bf\cdot}{\bf B}\right){\bf\cdot}d{\bf S}=0, \label{eq:2.2ba}
\end{equation}
where ${\bf B}{\bf\cdot}d{\bf S}$ is the magnetic flux through the area element $d{\bf S}$
which is advected with the flow.  
  Faraday's 
equation (\ref{eq:2.2ba})  can be described using the Calculus of exterior differential forms 
(e.g. \cite{Tur93},\cite{Webb14a}). It expresses the fact that
the magnetic flux ${\bf B}{\bf\cdot}d{\bf S}$ is conserved moving with the background flow. 
Note that $\nabla{\bf\cdot}{\bf B}=0$ (Gauss's law) is enforced by the 
Lagrange multiplier $\nu$.  

\cite{Yoshida09} studied the Clebsch expansion ${\bf u}=\nabla\phi+\alpha\nabla\beta$ 
and the completeness of the expansion. He showed that the Clebsch expansion:
\begin{equation}
{\bf u}=\nabla\phi+\sum_{j=1}^N \alpha_j \nabla\beta^j, \label{eq:gauge2.3}
\end{equation}
is in general complete if $N=n-1$ where $n$ is the number of independent variables 
 (i.e. the number of independent space variables).  
In some cases, it is necessary to control the boundary values of $\phi$, $\alpha_j$, and 
$\beta^j$, then $N=n$ (see e.g. \cite{Tanehashi15} 
who use Clebsch variables for both ${\bf u}$ and ${\bf B}$ in MHD gauge theory).
 In the remainder of the paper we use the Einstein summation convention for repeated indices.

The variational equations obtained by varying $\phi$,$\beta$, and $\lambda^k$
 and $\nu$ gives the equations:
\begin{align}
\frac{\delta{\cal A}}{\delta\phi}=&\rho_t+\nabla{\bf\cdot}(\rho {\bf u})=0,\quad 
\frac{\delta{\cal A}}{\delta\beta}= S_t+{\bf u}{\bf\cdot}\nabla S=0, \nonumber\\
\frac{\delta{\cal A}}{\delta\lambda^k}=&\frac{d\mu^k}{dt}=0, \quad 
\frac{\delta{\cal A}}{\delta\nu}=\nabla{\bf\cdot}{\bf B}=0, \label{eq:gauge2.4}
\end{align}
Varying $\boldsymbol{\Gamma}$ gives Faraday's equation (\ref{eq:gauge2.2a}). 
Equations (\ref{eq:gauge2.2a}) and (\ref{eq:gauge2.4}) are the constraint equations.

Varying ${\bf u}$ in the variational principle 
(\ref{eq:gauge2.2}) gives the Clebsch representation for ${\bf u}$ as:
\begin{equation}
{\bf u}=\nabla\phi-r\nabla S-\tilde{\lambda}^k\nabla\mu^k
+{\bf b}\times (\nabla\times{\boldsymbol\Gamma})
-\epsilon_1 \boldsymbol{\Gamma}\frac{\nabla{\bf\cdot}{\bf B}}{\rho}
, \label{eq:gauge2.5}
\end{equation}
where
\begin{equation}
r=\frac{\beta}{\rho},\quad \tilde{\lambda}^k=\frac{\lambda^k}{\rho},
\quad {\bf b}=\frac{\bf B}{\rho}.  \label{eq:gauge2.6}
\end{equation}

By varying ${\bf B}$ in the action principle gives:
\begin{equation}
\frac{\delta{\cal A}}{\delta\bf{B}}=-\biggl(\deriv{\boldsymbol\Gamma}{t}
- {\bf u}\times(\nabla\times{\boldsymbol\Gamma})
+\nabla(\nu+\epsilon_1 \boldsymbol{\Gamma}{\bf\cdot}{\bf u})
+\frac{\bf B}{\mu}\biggr)=0, \label{eq:gauge2.7} 
\end{equation}
for the evolution of ${\boldsymbol\Gamma}$. 

Varying $S$ gives the equation:
\begin{equation}
\frac{\delta{\cal A}}{\delta S}=-\left[\deriv{\beta}{t}+\nabla{\bf\cdot}(\beta {\bf u})+\rho T\right]=0
\quad\hbox{or}\quad \frac{dr}{dt}=-T, \label{eq:gauge2.8}
\end{equation}
where $T$ is the temperature of the gas and $r=\beta/\rho$. 
Here $d/dt=\partial/\partial t+{\bf u}{\bf\cdot}\nabla$ is the Lagrangian time derivative 
following the flow. 

Varying $\mu^k$ gives:
\begin{equation}
\frac{\delta{\cal A}}{\delta\mu^k}=
-\left\{\deriv{\lambda^k}{t}+\nabla{\bf\cdot} (\lambda^k {\bf u})\right\}=0\quad
\hbox{or}\quad 
\frac{d{\tilde\lambda}^k}{dt}=0, \label{eq:gauge2.9}
\end{equation}
where
\begin{equation}
{\tilde\lambda}^k=\frac{\lambda^k}{\rho}. \label{eq:gauge2.10}
\end{equation}

Varying $\rho$ results in Bernoulli's equation: 
\begin{equation}
\frac{\delta{\cal A}}{\delta\rho}==-\left\{\frac{d\phi}{dt}
-\left[\frac{1}{2} u^2-h-\Phi({\bf x})\right]\right\}=0, \label{eq:gauge2.11}
\end{equation}
where $\phi$ is the velocity potential,   
$h=\varepsilon_\rho=(p+\varepsilon)/\rho$ is the enthalpy of the gas and $p$ is the gas pressure. 

By taking the curl of (\ref{eq:gauge2.7}), we obtain: 
\begin{equation}
\deriv{\tilde{\boldsymbol\Gamma}}{t}-\nabla\times({\bf u}\times\tilde{\boldsymbol\Gamma})=-{\bf J}
\quad\hbox{where}\quad {\bf J}=\frac{\nabla\times{\bf B}}{\mu}
\quad\hbox{and}\quad \tilde{\boldsymbol\Gamma}=\nabla\times{\boldsymbol\Gamma}. 
\label{eq:gauge2.12}
\end{equation}
Equation({\ref{eq:gauge2.12}) has the same form as (\ref{eq:gauge1.2}) 
giving the current ${\bf J}$ in terms of the polarization ${\bf P}$ for the case 
$\nabla{\bf\cdot}{\bf P}=0$ in which ${\bf P}\to-\tilde{\boldsymbol\Gamma}$, and $\rho_c=0$. 
Thus, we identify ${\bf P}=-\tilde{\boldsymbol\Gamma}=-\nabla\times\boldsymbol{\Gamma}$.

If one uses the Lagrangian density 
\begin{align}
\ell_2=&\left(\frac{1}{2}\rho u^2-\varepsilon(\rho,S)-\frac{B^2}{2\mu}
-\rho\Phi({\bf x})\right) 
+\phi\left(\deriv{\rho}{t}+\nabla{\bf\cdot}(\rho {\bf u})\right)\nonumber\\
&+\beta\left(\deriv{S}{t}+{\bf u}{\bf\cdot}\nabla S\right)+  \lambda^k 
\left(\deriv{\mu^k}{t}+{\bf u}{\bf\cdot}\nabla \mu^k\right)\nonumber\\
&+{\boldsymbol\gamma}{\bf\cdot}\left(\deriv{\bf A}{t}
-{\bf u}\times(\nabla\times{\bf A})+\nabla({\bf u}{\bf\cdot}{\bf A})\right), 
\label{eq:gauge2.13}
\end{align}
to replace $\ell$ in the action (\ref{eq:gauge2.1}), then the equation:
\begin{equation}
\deriv{\bf A}{t}-{\bf u}\times(\nabla\times{\bf A})+\nabla({\bf u}{\bf\cdot A})=0, 
\label{eq:gauge2.14}
\end{equation}
replaces the Faraday equation constraint. It corresponds to using a gauge in which 
 the electric field potential $\psi$ is given by  $\psi={\bf u\cdot A}$. Here the one-form 
$\alpha={\bf A}{\bf\cdot}d{\bf x}$ 
is Lie dragged with the background flow. The curl of (\ref{eq:gauge2.14}) 
gives Faraday's equation. The action ${\cal A}_2$ corresponding to $\ell_2$ 
has Clebsch expansion for ${\bf u}$ of the form:
\begin{equation}
{\bf u}=\nabla\phi-r\nabla S- \tilde{\lambda}^k\nabla\mu^k+{\bf b}\times{\boldsymbol\gamma} 
+\nabla{\bf\cdot}{\boldsymbol\gamma}\ \frac{\bf A}{\rho}, \label{eq:gauge2.15}
\end{equation}
 where $\phi$ is the velocity potential.  

By varying ${\bf A}$ in ${\cal A}_2$ we obtain the equation:
\begin{equation}
\deriv{\boldsymbol\gamma}{t}-\nabla\times({\bf u}\times{\boldsymbol\gamma})
+{\bf u}\nabla{\bf\cdot}
{\boldsymbol\gamma}=-{\bf J}\equiv -\frac{\nabla\times{\bf B}}{\mu}. \label{eq:gauge2.16}
\end{equation}
In this approach, we identify ${\bf P}=-{\boldsymbol\gamma}$. 

\subsection{The MHD momentum equation}

The momentum equation for the fluid 
arises from the variational equations (\ref{eq:gauge2.4})-(\ref{eq:gauge2.12}).
The analysis is carried out for the case $\epsilon_1=1$ (i.e. for the case where 
$\nabla{\bf\cdot}{\bf B}\neq 0$). The results also apply for the case $\epsilon_1=0$
and $\Theta=\nabla{\bf\cdot}{\bf B}/\rho=0$. 
The MHD momentum equation is:
\begin{equation}
\derv{t}(\rho {\bf u})+\nabla{\bf\cdot}\biggl(\rho{\bf u}\otimes{\bf u}
+\left(p+\frac{B^2}{2\mu_0}\right) {\sf I} -\frac{{\bf B}\otimes{\bf B}}{ 2\mu_0}\biggr)
=-\rho \nabla\Phi, \label{eq:gauge2.16a}
\end{equation}
where $\Phi({\bf x})$ is the external gravitational potential. 
By using the mass continuity
equation, (\ref{eq:gauge2.16a}) reduces to the equation:
\begin{equation}
\frac{d{\bf u}}{dt}=-\frac{1}{\rho}\nabla p+\frac{{\bf J}\times {\bf B}}{\rho}
+{\bf B}\frac{\nabla{\bf\cdot B}}{\mu_0\rho}
-\nabla\Phi. \label{eq:gauge2.17}
\end{equation}

By using the identities:
\begin{equation}
-\frac{1}{\rho}\nabla p=T\nabla S-\nabla h,\quad 
{\bf u\cdot}\nabla {\bf u}=-{\bf u}\times\boldsymbol{\omega}
+\nabla\left(\frac{1}{2} u^2\right), \label{eq:gauge2.18}
\end{equation}
where $\boldsymbol{\omega}=\nabla\times{\bf u}$ is the fluid vorticity 
and $h=\varepsilon_{\rho}(\rho,S)$ is the gas 
enthalpy, (\ref{eq:gauge2.17}) takes the form: 
\begin{equation}
\Delta\equiv \deriv{\bf u}{t}-{\bf u}\times \boldsymbol{\omega} +\nabla 
\left(h+\Phi+\frac{1}{2} u^2\right)-T\nabla S-{\bf J}\times{\bf b}
-{\bf B}\frac{\nabla{\bf\cdot}{\bf B}}{\mu_0\rho}=0. \label{eq:gauge2.19}
\end{equation}
For the case $\epsilon_1=1$, ${\bf b}={\bf B}/\rho$ is an invariant vector field that is 
Lie dragged 
by the flow, i.e.,  
\begin{equation}
\left(\derv{t}+{\cal L}_{\bf u}\right) {\bf b}=\deriv{\bf b}{t}+[{\bf u},{\bf b}]\equiv
\deriv{\bf b}{t}+ {\bf u\cdot}\nabla{\bf b}-{\bf b\cdot}\nabla {\bf u}=0. 
\label{eq:gauge2.20}
\end{equation}
Equation (\ref{eq:gauge2.20}) is equivalent to Faraday's equation,  
 taking into account the mass continuity equation. The result (\ref{eq:gauge2.20}) 
holds for  $\nabla{\bf\cdot}{\bf B}=0$ and for the case 
$\nabla{\bf\cdot}{\bf B}\neq 0$.  
 ${\cal L}_{\bf u}{\bf b}=[{\bf u},{\bf b}]$ is the Lie derivative along  
${\bf u}$ of  the vector field ${\bf b}$, and $[{\bf u},{\bf b}]$   
is the left Lie bracket of ${\bf u}$ and ${\bf b}$.

In Appendix A, we show how the momentum equation (\ref{eq:gauge2.19}) arises from the 
variational equations (\ref{eq:gauge2.4})-(\ref{eq:gauge2.12}).  
In physical applications $\nabla{\bf\cdot}{\bf B}=0$. 
In the 
non-canonical Poisson bracket (e.g. \cite{MorrisonGreene80, MorrisonGreene82}, 
\cite{Chandre13}), the effect of $\nabla{\bf\cdot}{\bf B}\neq 0$, 
has been investigated for theoretical reasons (see e.g. \cite{Squire13} for 
a discussion in equation (13) et seq.). The effect of $\nabla{\bf\cdot}{\bf B}\neq 0$
is  important in numerical MHD codes, where the effect of numerically generated 
$\nabla{\bf\cdot}{\bf B}\neq 0$ needs to be minimized in order to produce accurate 
solutions (e.g. \cite{Powell99}, \cite{Janhunen00}, \cite{Dedner02}, 
\cite{Webb10b}). \cite{Evans88}, \cite{Balsara04}, \cite{BalsaraKim04}, and \cite{Stone09} 
use a staggered grid approach in which the magnetic flux is calcuated on the 
faces of the computational cell in order to minimize $\nabla{\bf \cdot}{\bf B}\neq 0$ 
errors.  

\section{Gauge symmetries} 
In this section we study gauge symmetries for MHD for the case $\nabla{\bf\cdot}{\bf B}=0$ 
(i.e. $\epsilon_1=0$ in the Lagrangian variational principle 
(\ref{eq:gauge2.1})-(\ref{eq:gauge2.2})). 
Gauge symmetries in fluid dynamics and MHD are similar to fluid relabeling symmetries. 
Both these two types of symmetries do not change the `physical variables', e.g. 
$\rho$, ${\bf u}$, ${\bf B}$, $p$, and $S$, but they do allow for the hidden variables 
behind the scenes to change. Thus, for both gauge symmetries and for fluid relabeling 
symmetries, the Eulerian variations are zero, i.e.  
\begin{equation}
\delta\rho=\delta{\bf u}=\delta{\bf B}=\delta p=\delta S=0. \label{eq:gauge3.1}
\end{equation}
For both these symmetries, the change in the action should be zero, modulo a pure 
divergence term to order $\epsilon$, where $\epsilon$ is the infinitesimal version 
of the group parameter describing the deviation from the identity transformation.
 Note that the Euler Lagrange equations are invariant under the addition of 
a pure divergence term to the Lagrangian.  
Lie symmetries are important in Noether's 
theorems, in which conservation laws are related to symmetries of the action. 
In the present analysis, the variables
$M=\left\{\phi,\beta,\lambda^k,\mu^k,\nu,{\boldsymbol\Gamma}\right\}$ are allowed to change. 
If there are continuous functions defining the transformations, then Noether's 
second theorem is used to obtain conservation laws. 
  Lie point symmetries
of the MHD equations and combinations of the scaling symmetries give 
rise to conservation laws (e.g. \cite{Webb07}).  
Potential symmetries of the equations can give rise 
to nonlocal conservation laws (e.g. \cite{Akhatov91}, \cite{Bluman10}).
 We restrict our attention to gauge symmetries.

Consider the gauge symmetries for the electric (${\bf E}$) and magnetic (${\bf B}$) 
fields in Maxwell's equations and in MHD. In particular, Gauss's equation, and Faraday's equation:
\begin{equation}
\nabla{\bf\cdot}{\bf B}=0,\quad \nabla\times{\bf E}+{\bf B}_t=0, \label{eq:gauge3.2}
\end{equation}
are satisfied by choosing
\begin{equation}
\bf{B}=\nabla\times{\bf A},\quad {\bf E}=-{\bf A}_t-\nabla\psi, \label{eq:gauge3.3}
\end{equation}
where ${\bf A}$ is the magnetic vector potential and $\psi$ is the electrostatic potential.
Equations (\ref{eq:gauge3.2})-(\ref{eq:gauge3.3}) remain invariant under the gauge 
\begin{equation}
{\bf A}'={\bf A}+\nabla\lambda\quad\hbox{and}\quad \psi'=\psi-\lambda_t. \label{eq:gauge3.4}
\end{equation}
The transformations (\ref{eq:gauge3.4}) leave ${\bf E}$ and ${\bf B}$ invariant., i.e.,
\begin{equation}
{\bf E}'={\bf E},\quad \hbox{and}\quad {\bf B}'={\bf B}. \label{eq:gauge3.5}
\end{equation}
The gauge transformations (\ref{eq:gauge3.4}) in classical electromagnetism are associated 
with the charge conservation law (e.g. \cite{Calkin63}). 

In the following analysis, we use the notation:
\begin{equation}
\delta\alpha=\epsilon V^\alpha, \label{eq:gauge3.5a}
\end{equation}
relating the variable $\alpha$ to its infinitesimal Lie group (or Lie pseudo group) generator
$V^\alpha$. 

For gauge symmetry transformations, 
 the physical variables (i.e. $(\rho,{\bf u},S,{\bf B},{\bf J})$) do not change. 
However, the   
 Lagrange multipliers and $\mu^k$ in general change under gauge transformations. 
From (\ref{eq:gauge2.5}) 
\begin{equation}
\delta{\bf u}=\epsilon V^{\bf u}=\epsilon\biggl[\nabla V^{\phi}-V^r\nabla S
-\left(V^{\tilde{\lambda}^k}\nabla\mu^k+\tilde{\lambda}^k \nabla(V^{\mu^k})\right)
+{\bf b}\times(\nabla\times V^{\boldsymbol\Gamma})\biggr]=0. \label{eq:gauge3.7}
\end{equation}
From (\ref{eq:gauge2.7}), the condition $\delta{\bf B}=0$ implies:
\begin{equation}
\deriv{V^{\boldsymbol\Gamma}}{t}-{\bf u}\times[\nabla\times(V^{\boldsymbol\Gamma})]
+\nabla V^\nu=0. \label{eq:gauge3.8}
\end{equation}
Equations (\ref{eq:gauge2.4}), (\ref{eq:gauge2.9}), (\ref{eq:gauge2.8}) 
and (\ref{eq:gauge2.11}) require:
\begin{equation}
\frac{d}{dt}V^{\mu^k}=\frac{d}{dt}V^{\tilde{\lambda}^k}=0, 
\quad \frac{d}{dt}V^r
=\frac{d}{dt}V^\phi=0. \label{eq:gauge3.9}
\end{equation}
Equations (\ref{eq:gauge3.9}) require $V^{\mu^k}$, $V^{\tilde{\lambda}^k}$, $V^r$ 
and $V^{\phi}$ to be functions only of the Lagrange labels.


The condition (\ref{eq:gauge3.7}) for $\delta{\bf u}=0$ 
can be reduced to a simpler form by introducing: 
\begin{equation}
G=\epsilon K= \delta\phi-\tilde{\lambda}^k\delta\mu^k \equiv \epsilon\left[ V^\phi
-\tilde{\lambda}^k V^{\mu^k}\right], \label{eq:gauge3.12}
\end{equation}
in (\ref{eq:gauge3.7}) (this definition of $G$ is like a Legendre transformation).
We obtain:
\begin{equation}
\nabla K=\nabla(V^\phi)-(\nabla\tilde{\lambda}^k)V^{\mu^k}
-\tilde{\lambda}^k\nabla V^{\mu^k}. \label{eq:gauge3.13}
\end{equation}
Re-arranging (\ref{eq:gauge3.13}) gives:
\begin{equation}
\nabla V^\phi-\tilde{\lambda}^k \nabla V^{\mu^k}=\nabla K
+\nabla\left(\tilde{\lambda}^k\right) V^{\mu^k}, \label{eq:gauge3.14}
\end{equation}
Using (\ref{eq:gauge3.14}),  (\ref{eq:gauge3.7}) reduces to:
\begin{equation}
\nabla K+\nabla(\tilde{\lambda}^k) V^{\mu^k}-V^{\tilde{\lambda}^k}\nabla\mu^k 
-V^r\nabla S+{\bf b}\times(\nabla\times V^{\boldsymbol{\Gamma}})=0 . \label{eq:gauge3.17}
\end{equation}
From (\ref{eq:gauge3.12}) 
\begin{equation}
V^\phi=K+\tilde{\lambda}^k V^{\mu^k}. \label{eq:gauge3.18}
\end{equation}
Using (\ref{eq:gauge3.18}) and (\ref{eq:gauge3.9}) it follows that:
\begin{equation}
\frac{dK}{dt}=0, \label{eq:gauge3.18a}
\end{equation}
(because $V^\phi$, $\tilde{\lambda}^k$ and $V^{\mu^k}$ are advected invariants). 
 In Appendix C, we show, that if $K=K(\tilde{\lambda},\mu,S)$ and ${\bf b}\times (\nabla\times V^{\boldsymbol\Gamma})=0$
then $K$ satisfies the Hamiltonian like equations: 
\begin{equation}
\frac{d\mu}{d\epsilon}=-\deriv{K}{\lambda},\quad \frac{d\lambda}{d\epsilon}=\deriv{K}{\mu},
\quad \frac{dr}{d\epsilon}=\deriv{K}{S}, 
\quad \frac{dS}{d\epsilon}=-\deriv{K}{r}=0, \label{eq:3.18aa}
\end{equation}
where $V^\alpha=d\alpha/d\epsilon$ is the Lie symmetry generator for the generic variable $\alpha$. 
This of course does not imply $K$ is the Hamiltonian for the system of 
determining equations (\ref{eq:gauge3.7})-(\ref{eq:gauge3.18a}).

Once the gauge field Lie determining equations 
(\ref{eq:gauge3.7})-(\ref{eq:gauge3.18a}) are solved 
for the Lie generators $(V^{\mu^k},V^{\tilde{\lambda}^k},V^\phi,V^r,V^{\boldsymbol{\Gamma}},
V^\nu)$, conservation laws for the extended MHD system,
can be obtained by using Noether's theorem. A form of Noether's 
theorem, suitable for this purpose is given below.

\section{Gauge symmetries and conservation laws}

In this section we use the gauge symmetries and Noether's theorem to 
derive conservation laws associated with the gauge symmetries of section 3. 
As in Section 3, we consider only the case where $\nabla{\bf\cdot}{\bf B}=0$ 
(i..e. $\epsilon_1=0$ in the variational principle (\ref{eq:gauge2.1})-(\ref{eq:gauge2.2})).

Starting from the action (\ref{eq:gauge2.2}), we look at the change in the action induced 
by infinitesimal  changes in the fields 
$\left\{\phi,\tilde{\lambda}^k,\mu^k, r, \nu, \boldsymbol{\Gamma}\right\}$. 
The change in the action is:
\begin{equation}
{\cal A}'-{\cal A}=\int\left(\ell'-\ell\right) \ d^3x dt, \label{eq:gauge4.1}
\end{equation}
where
\begin{align}
\ell'-\ell=&\delta\phi\left(\deriv{\rho}{t}+\nabla{\bf\cdot}(\rho{\bf u})\right) 
+\rho \left(\delta\tilde{\lambda}^k\frac{d\mu^k}{dt}
+\tilde{\lambda}^k \frac{d}{dt}\delta\mu^k\right)\nonumber\\ 
&+\rho\delta r\left(S_t+{\bf u}{\bf\cdot}\nabla S\right) +\delta{\boldsymbol\Gamma}
{\bf\cdot}\left(\deriv{\bf B}{t}-\nabla\times \left({\bf u}\times {\bf B}\right)\right)
+\delta\nu \nabla{\bf\cdot}{\bf B}+O\left(\epsilon^2\right). \label{eq:gauge4.2}
\end{align}
By noting that $d/dt (\delta\mu^k)=0$, 
and  integrating (\ref{eq:gauge4.2}) by parts, gives: 
\begin{align}
\ell'-\ell=&\derv{t}\left\{\rho\delta\phi+\rho S\delta r
+\rho\mu^k\delta \tilde{\lambda}^k 
+\delta{\boldsymbol\Gamma}{\bf\cdot}{\bf B}\right\}\nonumber\\
&+\nabla{\bf\cdot}\left\{ \rho {\bf u}\left(\delta\phi+S\delta r
+\mu^k\delta \tilde{\lambda}^k\right)
+{\bf B}\delta\nu +\delta\boldsymbol{\Gamma}\times({\bf u}\times {\bf B})\right\}\nonumber\\
&-\rho \frac{d}{dt}\delta\phi-S\rho \frac{d}{dt}\delta r
-\rho
\mu^k \frac{d}{dt}\delta\tilde{\lambda}^k \nonumber\\
&-\left(\mu^k\delta \tilde{\lambda}^k+S\delta r\right)
\left(\deriv{\rho}{t}+\nabla{\bf\cdot}(\rho{\bf u})\right)\nonumber\\
&-{\bf B}{\bf\cdot}\left(\deriv{\delta\boldsymbol{\Gamma}}{t}
-{\bf u}\times(\nabla\times\delta\boldsymbol{\Gamma})+\nabla(\delta\nu)\right). 
\label{eq:gauge4.3}
\end{align}

By using the gauge symmetry determining equations (\ref{eq:gauge3.8})-(\ref{eq:gauge3.9}) 
and noting that the non-divergence terms in (\ref{eq:gauge4.2}) vanish
because of the gauge symmetry equations (\ref{eq:gauge3.8})-(\ref{eq:gauge3.9}), 
the expression for $\ell'-\ell$ 
reduces to the sum of the pure time and space divergence terms. Using the notation
$\delta\alpha=\epsilon V^\alpha$ ($\alpha$ is a generic variable, that is varied 
in (\ref{eq:gauge4.1})-(\ref{eq:gauge4.3})), we obtain:
\begin{align}
\ell'-\ell=&\epsilon\biggl\{\derv{t}\left[\rho V^\phi+\rho S V^r
+\rho \mu^k V^{\tilde{\lambda}^k}
+{\bf V}^{\boldsymbol{\Gamma}}{\bf\cdot}{\bf B}\right]\nonumber\\
&+\nabla{\bf\cdot}\left[\rho{\bf u}\left(V^\phi+S V^r 
+\mu^k V^{\tilde{\lambda}^k}\right)
+{\bf B} V^\nu+ {\bf V}^{\boldsymbol\Gamma}\times({\bf u}\times{\bf B}\right]
\biggr\}\nonumber\\
&\equiv\epsilon\left(\deriv{W^0}{t}+\nabla{\bf\cdot}{\bf W}\right), \label{eq:gauge4.4}
\end{align}
where
\begin{align}
W^0=&\rho \left(V^\phi+ S V^r +\mu^k V^{\tilde{\lambda}^k}\right)
+{\bf V}^{\boldsymbol{\Gamma}}{\bf\cdot}{\bf B},\nonumber\\
{\bf W}=&\rho {\bf u}\left(V^\phi+S V^r 
+\mu^k V^{\tilde{\lambda}^k}\right)
+{\bf B} V^\nu+ {\bf V}^{\boldsymbol\Gamma}\times({\bf u}\times{\bf B}). \label{eq:gauge4.5}
\end{align}
One can add a pure divergence term to (\ref{eq:gauge4.4}) since it does not 
alter the Euler-Lagrange equations. Thus, more generally,
\begin{equation}
\delta \ell=\epsilon\left(\deriv{W^0}{t}
+\nabla{\bf\cdot}{\bf W}+\deriv{\Lambda^0}{t}+\nabla{\bf\cdot}
\boldsymbol{\Lambda}\right), \label{eq:gauge4.6}
\end{equation}
gives a more general form for the allowed changes in $\ell$, 
where $\Lambda^0$ and $\boldsymbol{\Lambda}$  are to be determined. For stationary variations, 
(\ref{eq:gauge4.6}), and for a finite transformation, not involving arbitrary functions,
(\ref{eq:gauge4.6}) gives the conservation law:
\begin{equation}
\derv{t}\left(W^0+\Lambda^0\right)
+\nabla{\bf\cdot}\left({\bf W}+\boldsymbol{\Lambda}\right)=0, 
\label{eq:gauge4.6a}
\end{equation}
which is Noether's first theorem for the case of a divergence symmetry of the action 
involving the gauge potential $\Lambda^0$ and flux $\boldsymbol{\Lambda}$. 

Below, we give examples of the use of (\ref{eq:gauge4.4})-(\ref{eq:gauge4.6a}) 
in which we derive (a) the magnetic helicity conservation law and (b)\ 
the cross-helicity conservation law, (c)\ the unmagnetized fluid helicity conservation 
equation and other examples.  The 
magnetic helicity and cross helicity conservation laws were  determined 
by \cite{Calkin63}  for an isobaric gas equation of state and 
for his modified, non-neutral, MHD type equations. The method used  
to obtain solutions of the Lie determining equations (\ref{eq:gauge3.8})-(\ref{eq:gauge3.17})
are outlined in Appendices B and  C.

\subsection{Magnetic helicity}
The magnetic helicity conservation law arises from the solution of 
(\ref{eq:gauge3.8})-(\ref{eq:gauge3.18a}), for which: 
\begin{equation}
{\bold V}^{\boldsymbol\Gamma} ={\bf A},\quad V^\nu=\psi,\quad K=0, 
\quad V^{\mu^k}=V^{\tilde{\lambda}^k}=V^\phi=V^r=0. \label{eq:gauge4.7}
\end{equation}
With this choice of parameters the conserved density $W^0$ and flux ${\bf W}$ 
in (\ref{eq:gauge4.5}) become:
\begin{align}
W^0=&{\bold V}^{\boldsymbol\Gamma}{\bf\cdot}{\bf B}={\bf A}{\bf\cdot}{\bf B}, \nonumber\\
{\bf W}=&{\bf B}\psi+{\bf A}\times({\bf u}\times{\bf B})
\equiv {\bf B}\psi+ ({\bf A}{\bf\cdot}{\bf B}){\bf u} -{\bf A}{\bf\cdot}{\bf u}){\bf B}\nonumber\\
=& ({\bf A}{\bf\cdot}{\bf B}){\bf u} +\left(\psi-{\bf A}{\bf\cdot}{\bf u}\right){\bf B}. 
\label{eq:gauge4.8}
\end{align}
The resultant conservation law, using Noether's theorem (\ref{eq:gauge4.6a}) gives:
\begin{equation}
\derv{t}({\bf A}{\bf\cdot}{\bf B})
+\nabla{\bf\cdot}\left[({\bf A}{\bf\cdot}{\bf B}){\bf u} 
+\left(\psi-{\bf A}{\bf\cdot}{\bf u}\right){\bf B}\right]=0. \label{eq:gauge4.9}
\end{equation}
which is the  Eulerian form of the magnetic helicity conservation law, where $h_M={\bf A\cdot B}$ 
is the magnetic helicity density. Integration of (\ref{eq:gauge4.9}) over a volume $V$ 
moving with the fluid, in which ${\bf B}{\bf\cdot n}=0$ on the boundary $\partial V$ of $V$ 
gives the Lagrangian magnetic helicity conservation law $dH_M/dt=0$, 
where $H_M$ is given by  (\ref{eq:gauge1.7}).  

\subsection{Cross helicity}
Using solutions of the Lie determining equations (\ref{eq:gauge3.8})-(\ref{eq:gauge3.18a})
( see Appendix B: set $k_2=k_1=1$, $\Lambda=\phi$ in (\ref{eq:B17})-(\ref{eq:B18})): 
\begin{align}
V^\phi=&K+\tilde{\lambda} {\bf b}{\bf\cdot}\nabla\mu,\quad \hbox{where}
\quad {\bf B}{\bf\cdot}\nabla K=0,\nonumber\\
V^\mu=&{\bf b}{\bf\cdot}\nabla\mu,\quad V^{\tilde\lambda}={\bf b}{\bf\cdot}\nabla\tilde{\lambda}, 
\quad V^r=0, \nonumber\\
V^{\boldsymbol\Gamma}=&\nabla\phi+\gamma\nabla K-\tilde{\lambda}\nabla\mu, 
\quad V^\nu=-\frac{d\phi}{dt} +{\bf u\cdot}V^{\boldsymbol\Gamma}, \label{eq:ch1}
\end{align}
in Noether's theorem (\ref{eq:gauge4.5}) gives:
\begin{equation}
W^0=\rho[K+{\bf b}{\bf\cdot}\nabla(\tilde{\lambda}\mu)]+h_c,\quad
{\bf W}={\bf u} W^0-\frac{d\phi}{dt} {\bf B},  \label{eq:ch2}
\end{equation}
where
\begin{equation}
h_c={\bf B}{\bf\cdot}({\bf u}+r\nabla S), \label{eq:ch3}
\end{equation}
is the generalized cross helicity density for non-barotropic flows (\cite{Webb14a, Webb14b}, 
Yahalom (2016)), and $\phi$ is the velocity potential in the Clebsch expansion for ${\bf u}$
which satisfies Bernoulli's equation (\ref{eq:gauge2.11}). We obtain:
\begin{equation}
\deriv{W^0}{t}+\nabla{\bf\cdot}{\bf W}=\deriv{h_c}{t}+\nabla{\bf\cdot}\left[{\bf u}h_c
+{\bf B}\left(h+\Phi-\frac{1}{2} u^2\right)\right]=0, \label{eq:ch4}
\end{equation}
which is the generalized cross helicity conservation law derived by \cite{Webb14a, Webb14b}
(an equivalent form of this conservation law is derived by \cite{Yahalom16},  see also Appendix E). 
In the derivation of (\ref{eq:ch4}) we have used the conservation law:
\begin{equation}
\deriv{(\rho \Lambda^0)}{t}+\nabla{\bf\cdot}(\rho {\bf u}\Lambda^0)
=\rho \frac{d\Lambda^0}{dt}=0\quad 
\hbox{where}\quad \Lambda^0=K+{\bf b}{\bf\cdot}\nabla(\tilde{\lambda}\mu). \label{eq:ch5}
\end{equation}
Note that $dK/dt=0$ and $d({\bf b}{\bf\cdot}\nabla(\tilde{\lambda}\mu))/dt=0$ in (\ref{eq:ch5}).

Equation(\ref{eq:ch4}) may be written in the more explicit form:
\begin{equation}
\derv{t}\left[{\bf B}{\bf\cdot}({\bf u}+r\nabla S)\right]
+\nabla{\bf\cdot}\left[{\bf B}{\bf\cdot}({\bf u}+r\nabla S){\bf u}
+{\bf B}\left(h+\Phi-\frac{1}{2}u^2\right)\right]=0, 
\label{eq:ch6}
\end{equation}
which is the generalized cross helicity conservation law for non-barotropic flows, in which
$p=p(\rho,S)$. Equation (\ref{eq:ch6}) is a nonlocal conservation law, because:
\begin{equation}
\frac{dr}{dt}\equiv\left(\deriv{r}{t}+{\bf u}{\bf\cdot}\nabla r\right)=-T, 
\label{eq:ch7}
\end{equation}
where $T$ is the temperature of the gas.

 Integration of (\ref{eq:ch6}) over a volume $V$ 
moving with the fluid for which ${\bf B\cdot n}=0$ on the boundary $\partial V$ of $V$, 
gives the generalized non-barotropic cross helicity conservation law:
\begin{equation}
\frac{dH_{CNB}}{dt}=0\quad \hbox{where}\quad H_{CNB}=\int_V \left({\bf u}+r\nabla S\right){\bf\cdot} {\bf B}\ d^3x, 
\label{eq:ch7a}
\end{equation}
(see also Appendix E and \cite{Yahalom17a, Yahalom17b}).

\subsection{The gauge symmetry ${\bf V}^{\boldsymbol\Gamma}=\nabla\Lambda$ and $V^\nu=-\Lambda_t$}
 
Set $k_2=1$, $k_1=0$, $K=0$ in (\ref{eq:B17})-(\ref{eq:B18}) in  Appendix B. 
Using the results in Appendix B, we obtain solutions of (\ref{eq:gauge3.7})-(\ref{eq:gauge3.18a}) 
of the form:
\begin{equation}
{\bf V}^{\boldsymbol\Gamma}=\nabla\Lambda,\quad V^\nu=-\Lambda_t, \quad
V^{\mu^k}=V^{{\tilde\lambda}^k}=V^r=V^\phi=K=0. \label{eq:gauge4.26}
\end{equation}
Use of Noether's theorem (\ref{eq:gauge4.6a}) gives the conservation law:
\begin{equation}
\derv{t}({\bf B}{\bf\cdot}\nabla\Lambda)+\nabla{\bf\cdot}
\left[-\Lambda_t {\bf B} +\nabla\Lambda\times({\bf u}\times{\bf B})\right]=0. \label{eq:gauge4.28}\end{equation}
This conservation law holds for all potentials $\Lambda({\bf x},t)$, where $\Lambda({\bf x},t)$ 
is not necessarily related to the MHD equations. One might regard (\ref{eq:gauge4.28}) 
as a trivial conservation law. However, if $\Lambda$ is related to the MHD equations, it does give
rise to interesting conservation laws. The conservation law (\ref{eq:gauge4.28}) can be written 
in the form:
\begin{equation}
\derv{t}({\bf B}{\bf\cdot}\nabla\Lambda)+\nabla{\bf\cdot}
\biggl[-(\Lambda_t+{\bf u\cdot}\nabla\Lambda){\bf B}
+({\bf B\cdot}\nabla\Lambda) {\bf u}\bigg]=0. \label{eq:gauge4.29}
\end{equation}
Some examples of the use of (\ref{eq:gauge4.29}) are discussed below.

\leftline{\bf Example 1}

If $\Lambda$ is advected with the flow, then $d\Lambda/dt=0$. 
In this case (\ref{eq:gauge4.29}) reduces to:
\begin{equation}
\derv{t}({\bf B}{\bf\cdot}\nabla\Lambda)
+\nabla{\bf\cdot}[({\bf B\cdot}\nabla\Lambda) {\bf u}]=0. \label{eq:gauge4.31}
\end{equation}
Thus, if $\Lambda=S$ then 
\begin{equation}
\derv{t}({\bf B}{\bf\cdot}\nabla S) 
+\nabla{\bf\cdot}\left[({\bf B}{\bf\cdot}\nabla S){\bf u}\right]=0. 
\label{eq:4.32}
\end{equation}
There are many examples of physically significant scalars advected with the flow. 
For example
\begin{equation}
\frac{d}{dt}\left(\frac{\bf A\cdot B}{\rho}\right)=0\quad \hbox{if}\quad 
\psi={\bf A\cdot u}\quad\hbox{and}\quad {\bf E}=-{\bf A}_t-\nabla\psi=-{\bf u}\times{\bf B}. 
\label{eq:gauge4.33}
\end{equation}
Thus, the choice $\Lambda={\bf A\cdot B}/\rho$ satisfies $d\Lambda/dt=0$ 
and gives rise to a physically relevant conservation law of the form (\ref{eq:gauge4.31}). 

\leftline{\bf Example 2}
Setting $\Lambda=\phi$,  
 the conservation law (\ref{eq:gauge4.29}) becomes:
\begin{equation}
\derv{t}\left({\bf B\cdot}\nabla\phi\right)
+\nabla{\bf\cdot}\left[({\bf B\cdot}\nabla\phi){\bf u} -{\bf B} \frac{d\phi}{dt}\right]=0. 
\label{eq:gauge4.34}
\end{equation}
Using Bernoulli's equation (\ref{eq:gauge2.11}):
\begin{equation}
\frac{d\phi}{dt}=\frac{1}{2}u^2-h-\Phi, \label{eq:4.35}
\end{equation}
in (\ref{eq:gauge4.34}) gives the conservation law:
\begin{equation}
\derv{t}\left({\bf B\cdot}\nabla\phi\right)
+\nabla{\bf\cdot}
\left[({\bf B\cdot}\nabla\phi){\bf u} +{\bf B}\left(h+\Phi-\frac{1}{2} u^2\right)\right] =0.
\label{eq:gauge4.35a}
\end{equation}
This conservation law is a nonlocal conservation law because 
\begin{equation}
\phi=\int_0^t \left(\frac{1}{2}u^2-h-\Phi\right) dt+\phi_0, \label{eq:gauge4.35b}
\end{equation}
is given by a Lagrangian time integral in which the memory of the flow plays a crucial role
($\phi_0=\phi({\bf x}_0,0)$ describes the initial data for the integral (\ref{eq:gauge4.35b})). 

Equation (\ref{eq:gauge4.28}) is a special case of a class of conservation laws 
for fluid systems obtained by \cite{Cheviakov14}. He showed that the system:  
\begin{equation}
\deriv{\bf N}{t}+\nabla\times {\bf M}=0\quad \hbox{and}\quad \nabla{\bf\cdot}{\bf N}=0, 
\label{eq:gauge4.36}
\end{equation}
has conservation laws of the form:
\begin{equation}
\derv{t}({\bf N}{\bf\cdot}\nabla F)
+\nabla{\bf\cdot}\left({\bf M}\times \nabla F-F_t{\bf N}\right)=0, \label{eq:gauge4.37}
\end{equation}
where $F({\bf x},t)$ is an arbitrary function of ${\bf x}$ and $t$, not necessarily related 
to the system (\ref{eq:gauge4.36}). 
In the MHD application (\ref{eq:gauge4.28}) to Faraday's equation, 
\begin{equation}
{\bf N}={\bf B},\quad {\bf M}=-({\bf u}\times{\bf B}),\quad F=\phi. \label{eq:gauge4.38}
\end{equation}

\cite{WebbMace15} using a fluid relabelling symmetry, and Noether's second theorem,
derived the conservation law:
\begin{equation}
\derv{t}\left(\boldsymbol{\omega}{\bf\cdot}\nabla\psi\right)
+\nabla{\bf\cdot}\biggl[\left(\boldsymbol{\omega}{\bf\cdot}\nabla\psi\right) {\bf u}
-\left(T\nabla S+\frac{{\bf J}\times{\bf B}}{\rho}\right)
\times\nabla\psi\biggr]=0, 
\label{eq:gauge4.39}
\end{equation}
where $\psi$ is a scalar advected with the flow, (i.e. $d\psi/dt=0$), 
 and $\boldsymbol{\omega}=\nabla\times{\bf u}$ is the fluid vorticity.  Here, 
${\bf J}=\nabla\times{\bf B}/\mu_0$ is the current and $T$ is the temperature of the 
gas. The conservation law (\ref{eq:gauge4.39}) corresponds to the choices:
\begin{align}
{\bf N}=&\boldsymbol{\omega}=\nabla\times{\bf u},\quad F=\psi({\bf x},t),\nonumber\\
{\bf M}=&-{\bf u}\times\boldsymbol{\omega}-\left(T\nabla S
+\frac{{\bf J}\times{\bf B}}{\rho}\right), \label{eq:gauge4.40}
\end{align}
in Cheviakov's potential vorticity type equation (\ref{eq:gauge4.37}).
\cite{Rosenhaus16} develop Noether's second theorem for quasi-Noether systems, 
and describe the conservation laws obtained by \cite{Cheviakov14} and 
\cite{CheviakovOberlack14}.

\subsection{Fluid helicity}

For an ideal, non-barotropic fluid, (${\bf B}=0$), the Clebsch expansion for ${\bf u}$ 
and related equations of use are:
\begin{align}
{\bf u}=&\nabla\phi-r\nabla S-\tilde{\lambda}^k\nabla\mu^k,
\quad {\bf w}={\bf u}+r\nabla S, \nonumber\\
\boldsymbol{\Omega}=&\nabla\times {\bf w}=-\nabla\tilde{\lambda}^k\times\nabla \mu^k
\equiv \boldsymbol{\omega}+\nabla r\times\nabla S,
\nonumber\\
\frac{dr}{dt}+T=&0,\quad \boldsymbol{\omega}=\nabla\times{\bf u},  \label{eq:FH!}
\end{align}
where $\boldsymbol{\omega}$ is the fluid vorticity. 
 The vorticity 
2-form $\boldsymbol{\Omega} {\bf\cdot} dS$ is Lie dragged with the flow, i.e.
\begin{equation}
\frac{d}{dt}\left(\boldsymbol{\Omega} {\bf\cdot} dS\right)
=\left(\derv{t}+{\cal L}_{\bf u}\right)\left(\boldsymbol{\Omega} {\bf\cdot} dS\right)
=\left[\deriv{\boldsymbol\Omega}{t}
-\nabla\times\left({\bf u}\times \boldsymbol{\Omega}\right) 
+\bf{u}\nabla{\bf\cdot}\boldsymbol{\Omega}\right]{\bf\cdot}dS=0, \label{eq:FH2}
\end{equation}
(e.g. \cite{Webb14a}). Note that:
\begin{equation}
\nabla{\bf\cdot}\boldsymbol{\Omega}=0. \label{eq:FH3}
\end{equation}
Equations (\ref{eq:FH2}) and (\ref{eq:FH3})  show that $\boldsymbol{\Omega}$ 
is analogous to ${\bf B}$ in MHD in Faraday;s equation, and in Gauss's equation 
$\nabla{\bf\cdot}{\bf B}=0$. Thus, using the analogy:
\begin{equation}
{\bf B}\to \boldsymbol{\Omega},\quad 
{\bf b}=\frac{\bf B}{\rho}\to \frac{\boldsymbol{\Omega}}{\rho}, 
\label{eq:FH4}
\end{equation}
it follows that the fluid helicity equation for non-barotropic fluids has the form:
\begin{equation}
\derv{t}\left[({\bf u}+r\nabla S){\bf\cdot}\boldsymbol{\Omega}\right]
+\nabla{\bf\cdot}\left[\left[({\bf u}+r\nabla S){\bf\cdot}\boldsymbol{\Omega}\right] {\bf u}
+\boldsymbol{\Omega}\left(h+\Phi-\frac{1}{2} u^2\right)\right]=0, \label{eq:FH5}
\end{equation}
(see e.g. \cite{Mobbs81} and \cite{Webb14a, Webb14b}
for vorticity theorems for non-barotropic fluids). 
Equation (\ref{eq:FH5}) is analogous to the 
cross helicity conservation law (\ref{eq:ch6}). In fact, one can derive the generalized 
fluid helicity conservation conservation law (\ref{eq:FH5}) 
by using the analysis of (\ref{eq:ch1}) seq. 
(see \cite{Webb14a, Webb14b} for alternative proofs of (\ref{eq:FH5})). 
For the case of an isobaric equation of state for the gas, (i.e. $p=p(\rho)$),
(\ref{eq:FH5}) reduces to the usual kinetic fluid  helicity conservation law:
\begin{equation}
\derv{t}({\bf u}{\bf\cdot}\boldsymbol{\omega})
+\nabla{\bf\cdot}\left[({\bf u}{\bf\cdot}\boldsymbol{\omega}){\bf u}
+\boldsymbol{\omega}\left(h+\Phi-\frac{1}{2} u^2\right)\right]=0. \label{eq:FH6}
\end{equation}
The fluid helicity conservation law (\ref{eq:FH5}) was derived by \cite{Webb14b},
by using a fluid relabelling symmetry

\subsection{Basic conservation laws}
Noether's theorem (\ref{eq:gauge4.5})-(\ref{eq:gauge4.6a}) 
  covers the basic MHD conservation laws. 
For example, setting $K=V^\phi=const.=c_1$ and all other symmetry generators in 
(\ref{eq:gauge4.5})-(\ref{eq:gauge4.6a}) equal to zero, gives the mass conservation 
law: $c_1[\rho_t+\nabla{\bf\cdot}(\rho {\bf u})]=0$. 
\cite{Holm98} describe the mass conservation law as being a consequence 
of symmetry breaking of the fluid relabelling symmetries [this interpretation 
is consistent with the above derivation, as $V^\phi=0$ gives no conservation law, 
but for $V^\phi=c_1$ gives the mass conservation law]. For $V^\nu=1$ and all other 
generators zero, in (\ref{eq:gauge4.5})-(\ref{eq:gauge4.6a}) 
 gives Gauss's law $\nabla{\bf\cdot}{\bf B}=0$. 
For $V^r=1$ and $K=S$, and all other generators zero, gives the entropy conservation 
law in the form: $2[(\rho S)_t+\nabla{\bf\cdot}(\rho S {\bf u})]=0$. 
For the choice $V^{\boldsymbol\Gamma}={\bf k}=const.$ and all other generators zero, 
gives Faraday's law in the form:
${\bf k}{\bf\cdot}[{\bf B}_t-\nabla\times({\bf u}\times{\bf B})]=0$. 

\section{Gauge Symmetries and Casimirs}

\cite{Henyey82} investigated the role of gauge symmetries in MHD 
using a Clebsch variable formulation of the equations. 
Henyey used the fact, that the Clebsch variable formulation 
yields canonical equations for the Hamiltonian, in which the physical 
variables $\rho$, ${\bf B}$ and $S$ can be regarded as canonical coordinates 
and the Lagrange multipliers 
are the corresponding canonical momenta. He considered gauge transformations 
in which the canonical coordinates ($\rho$,$S$,{\bf B}) are invariant, 
but the canonical momenta (the Lagrange constraint variables) are allowed 
to change.  \cite{Padhye96a, Padhye96b} showed that gauge 
transformations are related to the MHD Casimirs. \cite{Hameiri04} gives 
a thorough discussion of the MHD Casimirs. 
The Casimirs are functionals $C$ that have zero Poisson bracket with 
other functionals of the variables. 

\subsection{Henyey's approach}

In the symmetry group literature 
(e.g. \cite{Bluman89}), a gauge symmetry is sometimes referred to 
as a divergence symmetry, in which the action is invariant under a Lie 
transformation of the variables, and  involves a change 
in the Lagrangian density of the form ${\cal L}'={\cal L}+\epsilon 
\nabla_\alpha \Lambda^\alpha$.  
If the $\Lambda^\alpha=0$,  
 the symmetry is known as a variational symmetry. 

\cite{Henyey82}  does not enforce 
$\nabla{\bf\cdot}{\bf B}=0$, and omits the Lin constraint terms $\lambda^k d\mu^k/dt$ 
used in our analysis. He used  functionals $F$ of the physical variables 
$(\rho,S,{\bf B})$ which act as canonical coordinates $q^\alpha$,  and the Lagrange multipliers 
$(\phi,\beta,\boldsymbol{\Gamma})$ act as canonical momenta (see e.g. \cite{Zakharov97}). 
We include $\mu^k$ as canonical coordinates and the $\lambda^k$ 
as canonical momenta, and impose $\nabla{\bf\cdot}{\bf B}=0$ by  using  Lagrange mulipliers. 

For a finite dimensional Hamiltonian system, the change in the 
Lagrangian $\ell$, denoted by $\delta\ell=\ell'-\ell$ due to a canonical transformation
 corresponding to a gauge potential $F$ has the form:
\begin{equation}
\ell'-\ell=-\frac{dF}{dt}, \label{eq:gauge5.1}
\end{equation}
where $F$ is a functional of the canonical coordinates $(q^\alpha,p_\alpha)$. In the MHD case 
we set
\begin{equation}
F=F(\rho,S,{\bf B}, \mu^k; \phi,\beta,\boldsymbol{\Gamma}, \lambda^k). \label{eq:gauge5.2}
\end{equation}
In classical mechanics (e.g. \cite{Goldstein80}, ch. 9), the Lagrangian $\ell$ is related to the 
Hamiltonian $H({\bf q},{\bf p}, t)$ by the Legendre transformation:
\begin{equation}
\ell=p_k \dot{q}^k-H({\bf q},{\bf p}, t), \label{eq:gauge5.3}
\end{equation}
where we use the Einstein summation 
convention for repeated indicies $k$.  The Lagrangian in the new 
coordinates has the form:
\begin{equation}
\ell'=P_k \dot{Q}^k-K(Q^k,P_k,t), \label{eq:gauge5.4}
\end{equation}
where $K(Q^k,P_k,t)$ is the new Hamiltonian, (note $K$ in this section has a different meaning than that 
used in Section 4 ) and 
 $F$ is the generating function for the canonical transformation. For the  
 transformation (\ref{eq:gauge5.1})  the 
Euler Lagrange equations do not change under a divergence transformation (e.g. \cite{Bluman89} 
, \cite{Olver93}). If $F=F_1(q^k,Q^k,t)$ (\ref{eq:gauge5.3})-(\ref{eq:gauge5.4}) 
give the equation: 
\begin{equation}
p_k \dot{q}^k-H(q^k,p_k,t)=P_k \dot{Q}^k-K(Q^k,P_k,t)
+\left(\deriv{F_1}{t}+\deriv{F_1}{q^k}\dot{q}^k+ \deriv{F_1}{Q^k}\dot{Q}^k\right). 
\label{eq:gauge5.6}
\end{equation}
Collecting the the $\dot{q}^k$, $\dot{Q}^k$ and remaining terms in (\ref{eq:gauge5.6})
gives the canonical transformation equations:
\begin{equation}
p_k=\deriv{F_1}{q^k}, \quad P_k=-\deriv{F_1}{Q^k},\quad K=H+\deriv{F_1}{t}, 
\label{eq:gauge5.8}
\end{equation}
The gauge function $F_1(q^k,Q^k,t)$ defines the new canonical 
momentum variables $P_k$ in terms of the other variables (see \cite{Goldstein80}, Ch. 9 
for other 
possible choices of the gauge function $F$). 

\cite{Henyey82} considered gauge transformations in which the canonical coordinates 
$q^\alpha=Q^\alpha$  do not change (these are the physical 
variables $\rho$, $S$ and ${\bf B}$ and  the Lin constraint variables $\mu^k$), 
but the canonical momenta variables $(\phi,\beta,\boldsymbol{\Gamma},\lambda^k)$ 
do change.  Because  MHD is an infinite dimensional 
Hamiltonian system (e.g. \cite{MorrisonGreene80, MorrisonGreene82}, \cite{Holm83a}), 
it is necessary to use variational derivatives instead of partial 
derivatives in (\ref{eq:gauge5.8}). The MHD variational equations 
(\ref{eq:gauge2.4})-(\ref{eq:gauge2.12}) are invariant under the infinitesimal 
gauge transformations:
\begin{equation}
\left(\delta\phi,\delta\beta,\delta\boldsymbol{\Gamma}, \delta\lambda^k\right)
=\epsilon\left(F_\rho,F_S,F_{\bf B}, F_{\mu^k}\right), \quad
\left(\delta\rho,\delta S,\delta {\bf B},\delta\mu^k\right)=(0,0,0,0), \label{eq:gauge5.9}
\end{equation}
where $F=F (\rho,S,{\bf B},\mu^k)$ is the gauge function ($F=-F_1$ in the analogy 
(\ref{eq:gauge5.8})). Here we use the notation:
\begin{equation}
 \delta P_{\alpha}=P_\alpha-p_\alpha=\epsilon V^{P_\alpha}=\epsilon\frac{\delta F}{\delta Q^\alpha}
\equiv \epsilon F_{Q^\alpha}, \label{eq:gauge5.10}
\end{equation}
(we sometimes use $p'_\alpha\equiv P_\alpha$ to denote the transformed canonical momenta). 

In (\ref{eq:gauge2.1}), the fluid velocity is given 
by the Clebsch expansion:
\begin{equation}
{\bf u}=\nabla\phi-\frac{\beta}{\rho}\nabla S-\frac{\lambda^k}{\rho} \nabla\mu^k 
+\frac{{\bf B}\times(\nabla\times\boldsymbol{\Gamma})}{\rho}-
\frac{\boldsymbol{\Gamma}\nabla{\bf\cdot}{\bf B}}{\rho}. \label{eq:gauge5.11}
\end{equation}
The gauge transformation (\ref{eq:gauge5.9}) is required to leave ${\bf u}$ 
invariant to $O(\epsilon)$, i.e.
\begin{align}
\delta {\bf u}=&\epsilon\left\{\nabla V^\phi -\frac{V^\beta}{\rho}\nabla S
-\frac{V^{\lambda^k}}{\rho} \nabla\mu^k  
+\frac{{\bf B}\times\left(\nabla\times {\bf V}^{\boldsymbol\Gamma}\right)}{\rho}
-\frac{V^{\boldsymbol\Gamma}\nabla{\bf\cdot}{\bf B}}{\rho}\right\}
\nonumber\\
\equiv& \epsilon\left\{\nabla F_\rho-\frac{F_S}{\rho}\nabla S
-\frac{F_{\mu^k}}{\rho}\nabla\mu^k
+\frac{{\bf B}\times\left(\nabla\times F_{\bf B}\right)}{\rho}
-F_{\bf B}\frac{\nabla{\bf\cdot B}}{\rho}\right\} 
=0. \label{eq:gauge5.12}
\end{align}
There are further invariance conditions on the Euler Lagrange equations 
(\ref{eq:gauge2.4})-(\ref{eq:gauge2.11}) due to the gauge transformations, namely:
\begin{equation}
\frac{d}{dt} V^{\mu^k}=\frac{d}{dt} V^{\tilde{\lambda}^k}=0,
\quad \frac{d}{dt} V^r=\frac{d}{dt} V^\phi=0, \label{eq:gauge5.13}
\end{equation}
\begin{equation}
\derv{t} {\bf V}^{\boldsymbol\Gamma}
-{\bf u}\times\left(\nabla\times{\bf V}^{\boldsymbol\Gamma}\right)
+\nabla \left(V^\nu+V^{\boldsymbol\Gamma} {\bf\cdot}{\bf u}\right)=0. \label{eq:gauge5.14}
\end{equation}
These equations are the same as (\ref{eq:gauge3.8})-(\ref{eq:gauge3.9}).
  Note that:
\begin{equation}
V^{\tilde{\lambda}^k}=\frac{V^{\lambda^k}}{\rho},\quad V^r=\frac{V^\beta}{\rho}. 
\label{eq:gauge5.15}
\end{equation}
In Section 4, we allowed both $\lambda^k$ and 
$\mu^k$ to vary
(i.e. $V^{\lambda^k}\neq 0$ and/or $V^{\mu^k}\neq 0$ as possibilities, since both 
$\mu^k$ and $\lambda^k$ were not identified as physical variables). 
Equations (\ref{eq:gauge5.13}) may  be expressed in terms of the variational derivatives 
of $F$. 

Equation (\ref{eq:gauge5.12}) may be written as:
\begin{equation}
\rho \nabla F_\rho- F_{\mu^k} \nabla\mu^k-F_S \nabla S
+ {\bf B}\times\left(\nabla\times F_{\bf B}\right)
-F_{\bf B} \nabla{\bf\cdot}{\bf B}=0, \label{eq:gauge5.16}
\end{equation}
or in the form:
\begin{equation}
\rho \nabla F_\rho- F_{\mu^k} \nabla\mu^k- F_S \nabla S
+{\bf B}{\bf\cdot} \left(\nabla F_{\bf B}\right)^T-{\bf B\cdot}\nabla F_{\bf B}
-F_{\bf B} \nabla{\bf\cdot}{\bf B}=0, 
\label{eq:gauge5.17}
\end{equation}
which is equivalent to \cite{Henyey82}, equation (24) (the $\mu^k$ terms are 
not present in \cite{Henyey82}). By noting that
\begin{equation}
F_{\bf A}=\nabla\times F_{\bf B}. \label{eq:gauge5.18}
\end{equation}
 Note that ${\bf B}=\nabla\times{\bf A}$. (\ref{eq:gauge5.16}) may  be expressed as:
\begin{equation}
\rho \nabla F_\rho- F_{\mu^k} \nabla\mu^k-{F}_S \nabla S
+ {\bf B}\times F_{\bf A}-F_{\bf B} \nabla{\bf\cdot}{\bf B}=0, \label{eq:gauge5.19}
\end{equation}
which is useful in the case of magnetic helicity functionals. 

\begin{proposition}\label{prop5.1}
The  invariance condition (\ref{eq:gauge5.16}) in Henyey's (1982) gauge transformation 
can be written in the form: 
\begin{equation}
\rho\nabla{\bar F}_\rho - {\bar F}_S \nabla S+\boldsymbol{\omega}\times {\bar F}_{\bf u}
+{\bf B}\times (\nabla\times {\bar F}_{\bf B})- {\bar F}_{\bf B}\nabla{\bf\cdot}{\bf B}=0, 
\label{eq:gauge5.19d1}
\end{equation}
where $\boldsymbol{\omega}=\nabla\times{\bf u}$ is the fluid vorticity. The functional 
$\bar{F}(\rho,{\bf u},S,{\bf B})$ in (\ref{eq:gauge5.19d1}) is equivalent to (i.e. 
has the same value as) the functional 
$F=F(\rho,S,\mu,{\bf B}; \phi,\beta,\lambda,\boldsymbol{\Gamma})$. This result, 
coupled with the gauge transformations:
\begin{align}
\delta\rho=&F_\phi=-\nabla{\bf\cdot}\bar{F}_{\bf u}=0,
\quad \delta S=F_\beta=-\frac{\bar{F}_{\bf u}{\bf\cdot}\nabla S}{\rho}=0, \nonumber\\
\delta\mu=&F_{\lambda}=-\frac{{\bar F}_{\bf u}{\bf\cdot}\nabla\mu}{\rho}=0, \nonumber\\
\delta{\bf B}=&F_{\boldsymbol{\Gamma}}=\nabla\times(\bar{F}_{\bf M}\times {\bf B})
-\bar{F}_{\bf M}\nabla{\bf\cdot}{\bf B}=0, \label{eq:gauge5.19d2}
\end{align}
where
\begin{equation}
\bar{F}_{\bf M}=\frac{{\bar F}_{\bf u}}{\rho}, \label{eq:gauge5.19d3}
\end{equation}
 are the Casimir determining equations 
(cf \cite{Hameiri04}, \cite{Morrison98},  \cite{Padhye96a, Padhye96b}, 
\cite{Holm83a,Holm83b}). 
\end{proposition}
The detailed proof of Proposition \ref{prop5.1} is given in Appendix D.

\cite{Henyey82} observed that the gauge symmetry determining equation (\ref{eq:gauge5.16}) 
has solutions:
\begin{equation}
F=\int d^3x\ F\left(\nabla{\bf\cdot}{\bf B},{\bf x}\right) 
+\rho G\left(S,{\bf b\cdot}\nabla S, 
{\bf b\cdot}\nabla({\bf b\cdot}\nabla S),\ldots\right), \label{eq:gauge5.31a}
\end{equation}
where ${\bf b}={\bf B}/\rho$ (in Henyey's analysis $\nabla{\bf\cdot}{\bf B}=0$ is not imposed, 
but it is noted that if $\nabla{\bf\cdot}{\bf B}=0$ at time $t=0$ then 
$\nabla{\bf\cdot}{\bf B}=0$ for all $t>0$).


It was shown in Section 4, that the 
local magnetic helicity conservation  law (\ref{eq:gauge4.9}) 
arises in the Calkin approach by choosing ${\bf V}^{\boldsymbol\Gamma}={\bf A}$, 
$V^\nu=\psi$ and the other Lie symmetry generators in 
(\ref{eq:gauge4.7}) are set equal to zero in Noether's theorem. If one chooses the gauge of 
${\bf A}$ such that $\psi={\bf A\cdot u}$ then the one-form $\alpha={\bf A}{\bf\cdot}d{\bf x}$ is Lie dragged with the flow, and in that case 
the conservation law (\ref{eq:gauge4.9}) can be written in the form:
\begin{equation}
\frac{d}{dt}\left(\frac{\bf A\cdot B}{\rho}\right)=0. \label{eq:gauge5.24a}
\end{equation}
 \cite{Padhye96a, Padhye96b} give a class of 
MHD Casimir solutions of (\ref{eq:gauge5.19d1})-(\ref{eq:gauge5.19d3}) of the form:
\begin{equation}
C[\rho,S,{\bf A}]=\int_V\rho G\left({\bf A\cdot b}, {\bf b\cdot}\nabla S, 
{\bf b\cdot}\nabla({\bf b\cdot}\nabla S), {\bf b\cdot}\nabla({\bf A\cdot}{\bf b}),\ldots \right)
\ d^3x, \label{eq:gauge5.25a}
\end{equation}
for the gauge case where (\ref{eq:gauge5.24a}) applies (see e.g. \cite{Gordin87}, 
\cite{Gordin89}).

\section{Concluding Remarks}
In this paper we have explored the origin of conservation laws in MHD using the Clebsch gauge 
field theory approach of \cite{Calkin63}. One of the main motivations was to relate 
the charged fluid extended MHD model developed by \cite{Calkin63} to the more 
commonly used quasi-neutral, Clebsch approach to MHD (e.g. \cite{Zakharov97}; \cite{Holm83a}
; \cite{Morrison98}). 
A second motivation was to understand more clearly the gauge symmetry 
responsible for the magnetic helicity conservation law (\ref{eq:gauge4.9}) which 
does not arise as a fluid relabelling symmetry conservation law. 
In gauge transformations 
the physical variables $(\rho,{\bf u},{\bf B}, S)$ do not change, but the 
Lagrange multipliers and the Lin constraint variables are allowed to change. 

In \cite{Calkin63} the electric current ${\bf J}$ is expressed in terms of the polarization 
vector ${\bf P}$ (see (\ref{eq:gauge1.1})-(\ref{eq:gauge1.3}), 
in which the charge density $\rho_c$ is given by $\nabla{\bf\cdot}{\bf P}=-\rho_c$). 
Equation (\ref{eq:gauge1.6}) for ${\bf P}$ has the form of Faraday's equation for ${\bf P}$
in which the current ${\bf J}$ acts as a source term (see e.g. \cite{Panofsky64}).
We show that the curl of the Clebsch variable $\boldsymbol{\Gamma}$ 
 behaves like ${\bf P}$ 
(i.e. ${\bf P}=-\nabla\times\boldsymbol{\Gamma}$ in the case $\nabla{\bf\cdot}{\bf P}=0$). 
A similar result (equation (\ref{eq:gauge2.16})) may be obtained if Faraday's equation 
is expressed in terms of the magnetic vector potential ${\bf A}$. 

The Lie symmetry determining equations  for the Clebsch variables 
$\phi$, $r$, $\tilde{\lambda}^k$,$\mu^k$,$\nu$, and $\boldsymbol{\Gamma}$ 
follow from requiring 
the physical variables $(\rho,{\bf u},{\bf B}, S)$ to have zero variations under the 
transformations. Requiring the variation of the action to be zero to $O(\epsilon)$ 
to within a divergence transformation of the Lagrangian then gives Noether's theorem, 
which was used to obtain conservation laws for: (a)\ magnetic helicity, (b)\ cross helicity, 
(c)\  fluid helicity for a non-magnetized fluid and (d)\ a class of conservation 
laws associated with Faraday's equation. The latter conservation laws are a special case 
of conservation laws for curl and divergence systems of equations 
derived by \cite{Cheviakov14} (see e.g. \cite{Rosenhaus16} for an account 
involving Noether's second theorem for quasi-Noether systems, and \cite{WebbMace15} 
for a discussion of  potential vorticity type conservation laws in MHD). 

Section 5 extended the gauge transformation approach to MHD of \cite{Henyey82}. 
In this formulation, the physical variables $(\rho,S,{\bf B})$ and Lin constraint variables 
$\mu^k$ act as canonical coordinates, and the corresponding Lagrange multipliers 
$(\phi,\beta,\boldsymbol{\Gamma}, \lambda^k)$ correspond to canonical momenta. 
The canonical coordinates $(\rho,S,{\bf B},\mu^k)$ and the fluid velocity ${\bf u}$ 
 do not change, but the canonical momenta (the Lagrange multipliers) do change. 
We showed that the Henyey approach gives the Casimir determining equations derived 
by Hameiri (2004) and others.

The present approach can be expanded to take into account the MHD Lie point symmetries.
 \cite{Calkin63} used the space-time invariances of the action to derive 
momentum conservation equation, the energy conservation equation, and the angular 
momentum conservation equation, associated with space translation invariance, time translation
invariance and rotational invariance of the action (see e.g. \cite{Morrison82},
\cite{Webb05b} for the 10 Galilean Lie point symmetries).  
 \cite{Webb07, Webb09} noted that the scaling 
Lie point symmetries for special equations of state for the gas can be combined to 
give another set of conservation laws. \cite{Akhatov91}, \cite{Bluman10}, 
\cite{Sjoberg04} and \cite{Webb09} have derived nonlocal conservation 
laws associated with potential symmetries of the gas dynamic equations.  
The generalized helicity and cross helicity conservation 
laws for a non-barotropic gas correspond to nonlocal potential symmetries 
due to the Lagrange multiplier $r$ used to impose the entropy 
conservation equation (see also \cite{Mobbs81}). 
 
\cite{Yahalom13} discusses  magnetic helicity by using an analogy 
with the  
Aharonov-Bohm  (AB) effect in quantum mechanics.  This is related to 
the magnetic helicity and cross helicity of the flow.  
 The interpretation of cross helicity for non-barotropic flows 
and magnetic helicity as generalized AB effects is given in \cite{Yahalom16, Yahalom17a, Yahalom17b}, 
Appendix E).  
In \cite{Calkin63} the polarization ${\bf P}$ is used to 
describe the MHD variational principle, but in our approach the polarization ${\bf P}$ is 
a Lagrange multiplier enforcing Faraday's equation (see introduction).

The present analysis provides: (a)\ a direct derivation of 
the magnetic helicity conservation law using Noether's theorem and a gauge transformation 
symmetry (see e.g. \cite{Calkin63})  
 and (b)\ it provides a link between 
MHD and gauge field theories (e.g. \cite{Jackiw02, Jackiw04}, \cite{Kambe07, Kambe08}),
\cite{Banerjee16}). 
 \cite{Tanehashi15}, use the 
known Casimirs for barotropic MHD, and the non-canonical Poisson 
bracket of \cite{MorrisonGreene80, MorrisonGreene82} to 
uncover gauge symmetries in MHD, by using a Clebsch variable expansion for both 
${\bf u}$ and ${\bf B}$. 
\cite{Araki16} provides an alternative viewpoint of fluid relabelling symmetries in 
MHD involving generalized vorticity and normal mode expansions for ideal incompressible 
fluids and MHD by using integro-differential operators acting on the generalized 
velocities. 
 
The multi-symplectic approach to fluid dynamics has been explored by \cite{Hydon05}, 
\cite{Bridges05, Bridges10},  \cite{Cotter07}, \cite{Webb15}, and \cite{Webb16}. 
The exact connection of these approaches to the present approach 
remains to be explored.

\appendix
\section*{Appendix A}
\setcounter{section}{1}

The momentum equation (\ref{eq:gauge2.19}) is a consequence  of the variational equations
(\ref{eq:gauge2.4})-(\ref{eq:gauge2.12}). The right handside of the 
momentum equation (\ref{eq:gauge2.19}) may be written in the form:
\begin{align}
\Delta=&\nabla\left\{\frac{d\phi}{dt}-\left[\frac{1}{2} u^2-(h+\Phi)\right]\right\}
-\left(\frac{dr}{dt}+T\right) \nabla S\nonumber\\
&-r \nabla\left(\frac{dS}{dt}\right)
-\nabla\mu^k\frac{d\tilde{\lambda}^k}{dt}
-\tilde{\lambda}^k \nabla\left(\frac{d\mu^k}{dt}\right)\nonumber\\
&+{\bf b}\times\biggl\{\deriv{\tilde{\boldsymbol\Gamma}}{t}
-\nabla\times\left({\bf u}\times \tilde{\boldsymbol\Gamma}\right)+{\bf J}\biggr\}
+\left\{\deriv{\bf b}{t}+[{\bf u},{\bf b}]\right\}\times\tilde{\boldsymbol\Gamma}\nonumber\\
&-\Theta\left\{\deriv{\boldsymbol{\Gamma}}{t}-{\bf u}\times\tilde{\boldsymbol{\Gamma}}
+\nabla(\nu+\boldsymbol{\Gamma}{\bf\cdot u}) +\frac{\bf B}{\mu_0}\right\}
-\boldsymbol{\Gamma}\frac{d\Theta}{dt}+\Theta\nabla\nu\nonumber\\
&+\nabla\biggl\{{\bf u\cdot}\left[{\bf u}-\left(\nabla\phi-r\nabla S
-\tilde{\lambda}^k\nabla\mu^k+{\bf b}\times\tilde{\boldsymbol\Gamma}
-\boldsymbol{\Gamma}\Theta\right)\right]\biggr\}
+{\bf Q}, \label{eq:gauge2.21}
\end{align}
where 
\begin{equation}
\Theta=\nabla{\bf\cdot}{\bf B}/\rho, \label{eq:gauge2.21a}
\end{equation}
and 
\begin{align}
{\bf Q}=&{\bf b}\times\left[\nabla\times({\bf u}\times\tilde{\boldsymbol\Gamma})\right]
+{\bf u}\times\left[\nabla\times(\tilde{\boldsymbol\Gamma}\times{\bf b})\right]\nonumber\\
&+\tilde{\boldsymbol\Gamma}\times[{\bf u},{\bf b}]
+\nabla\left({\bf u\cdot}({\bf b}\times\tilde{\boldsymbol\Gamma})\right)= 0. 
\label{eq:gauge2.22}
\end{align}
One can verify that ${\bf Q}=0$ by collecting the derivatives of ${\bf u}$, ${\bf b}$ and 
$\tilde{\boldsymbol\Gamma}$ separately. An alternative proof that ${\bf Q}=0$ 
is an identity, may carried out by using the Calculus of exterior differential 
forms (see below). In (\ref{eq:gauge2.22}) we have dropped 
the $\nabla{\bf\cdot}\tilde{\boldsymbol\Gamma}$ terms because 
$\tilde{\boldsymbol\Gamma}=\nabla\times{\boldsymbol\Gamma}$ is a curl, 
and hence has zero divergence.

Note that by taking the divergence of the generalized Faraday equation (\ref{eq:gauge2.2a})
we obtain the conservation law
\begin{equation}
\derv{t}(\nabla{\bf\cdot}{\bf B})+\nabla{\bf\cdot}({\bf u}\nabla{\bf\cdot}{\bf B})=0
\quad \hbox{or}\quad \derv{t}(\rho\Theta) +\nabla{\bf\cdot}(\rho {\bf u}\Theta)
\equiv \rho \frac{d\Theta}{dt}=0. 
\label{eq:gauge2.22a}
\end{equation}
A variant of the identity (\ref{eq:gauge2.22}) 
was given by \cite{Calkin63} in establishing a generalized vorticity equation and a 
generalized momentum equation. 

A more symmetric way to write ${\bf Q}$ in (\ref{eq:gauge2.22}) is:
\begin{equation}
{\bf Q}=\sum \{{\bf b}\times[\nabla\times({\bf u}\times\tilde{\boldsymbol{\Gamma}})]
+({\bf u}\times{\bf b}) \nabla{\bf\cdot}\tilde{\boldsymbol{\Gamma}}
+\circlearrowright \}
+\nabla[{\bf u}{\bf\cdot}({\bf b}\times \tilde{\boldsymbol{\Gamma}})]=0
\label{eq:gauge2.22b}
\end{equation} 
where the symbol $\circlearrowright$  means cyclically 
permute $({\bf b}, {\bf u}, \tilde{\boldsymbol{\Gamma}})$ and then sum. It is interesting 
to note that $\nabla\times{\bf Q}=0$. 

Below, we derive the identity (\ref{eq:gauge2.22}) 
by using the algebra of exterior differential forms. The theory 
and formulae for the algebra of exterior differential forms was 
developed by Elie Cartan in his study of Lie symmetry transformations 
and their connection to differential geometry (see e.g., 
\cite{Marsden94}, \cite{Holm08a}). 
We use the results listed in \cite{Webb14a}) in our analysis. 
A basic formula in this theory is Cartan's magic formula, for the 
Lie derivative of a differential form $\omega$ with respect to a vector field 
${\bf V}$, namely:
\begin{equation}
{\cal L}_{\bf V} (\omega)={\bf V}\lrcorner d\omega+d({\bf V}\lrcorner\omega). 
\label{eq:A1}
\end{equation}
Here ${\cal L}_{\bf V}=d/d\epsilon$ denotes the Lie derivative with respect to 
a vector field ${\bf V}$, which is tangent to a curve with curve parameter 
$\epsilon$ (usually $\epsilon$ corresponds to the infinitesimal parameter 
of some set of curves associated with a Lie symmetry group). The quantity $d\omega$ 
is the exterior derivative of the differential form $\omega$. If $\omega$ is a 
$p$-form, then $d\omega$ is a $p+1$-form, and ${\cal L}_{\bf V}\omega$ 
is a $p$-form. The symbol $\wedge$ denotes the anti-symmetric 
wedge product and ${\bf V}\lrcorner \omega$ denotes the $(p-1)$-form 
that results from contracting the vector field ${\bf V}$ 
with the $p$-form $\omega$.  a related formula to (\ref{eq:A1}) 
is:
\begin{equation}
{\cal L}_{\bf V}\left({\bf u}\lrcorner\omega\right)={\cal L}_{\bf V}({\bf u})
\lrcorner \omega+{\bf u}\lrcorner {\cal L}_{\bf V}(\omega)
\equiv[{\bf V},{\bf u}]\lrcorner\omega+{\bf u}\lrcorner{\cal L}_{\bf V}(\omega)
\label{eq:A2}
\end{equation}
where ${\cal L}_{\bf V}=[{\bf V},{\bf u}]$ is the left Lie bracket of ${\bf V}$ and ${\bf u}$. 

To prove (\ref{eq:gauge2.22}), we start by using (\ref{eq:A2}), with ${\bf V}\to{\bf b}$, 
and $\omega\to \beta$ where
\begin{equation}
\beta=\tilde{\boldsymbol\Gamma}{\bf\cdot}d{\bf S}=\tilde{\Gamma}^x dy\wedge dz+
\tilde{\Gamma}^y dz\wedge dx+\tilde{\Gamma}^z dx\wedge dy,
\label{eq:A3}
\end{equation}
 ${\bf b}={\bf B}/\rho$, $\tilde{\boldsymbol\Gamma}=\nabla\times\boldsymbol{\Gamma}$ 
and ${\bf u}$ is the fluid velocity. We obtain:
\begin{equation}
{\cal L}_{\bf b}\left({\bf u}\lrcorner \beta\right)
= [{\bf b},{\bf u}]\lrcorner\beta+{\bf u}\lrcorner {\cal L}_{\bf b}(\beta). \label{eq:A4}
\end{equation}
Our strategy is to evaluate the individual terms in (\ref{eq:A4}) 
to obtain (\ref{eq:gauge2.22}). Using the formulas:
\begin{align}
{\bf u}\lrcorner\beta=&{\bf u}\lrcorner (\tilde{\boldsymbol\Gamma}{\bf\cdot} d{\bf S})
=-\left({\bf u}\times\tilde{\boldsymbol\Gamma}\right){\bf\cdot}d{\bf x}, \nonumber\\
[{\bf b},{\bf u}]\lrcorner\beta=&-[{\bf b},{\bf u}]\times \tilde{\boldsymbol\Gamma}
{\bf\cdot}d{\bf x}, \nonumber\\
{\cal L}_{\bf b}(\beta)=&{\cal L}_{\bf b}\left(\tilde{\boldsymbol\Gamma}
{\bf\cdot}d{\bf S}\right)=[-\nabla\times({\bf b}\times\tilde{\boldsymbol\Gamma})
+{\bf b}(\nabla{\bf\cdot}\tilde{\boldsymbol\Gamma})]{\bf\cdot}d{\bf S}, \label{eq:A5}
\end{align}
we obtain:
\begin{equation}
{\bf u}\lrcorner{\cal L}_{\bf b}(\beta)=-{\bf u}\times
\left\{-\nabla\times({\bf b}\times\tilde{\boldsymbol\Gamma})
+{\bf b}(\nabla{\bf\cdot}\tilde{\boldsymbol\Gamma})\right\} {\bf\cdot}d{\bf x}. \label{eq:A6}
\end{equation}

Next we evaluate ${\cal L}_{\bf b}({\bf u}\lrcorner \beta)$ in (\ref{eq:A4}). Writing
$\boldsymbol{\gamma}={\bf u}\lrcorner\beta$ and using Cartan's magic formula (\ref{eq:A1}) 
we obtain:
\begin{equation}
{\cal L}_{\bf b}({\bf u}\lrcorner\beta)\equiv{\cal L}_{\bf b}(\boldsymbol{\gamma})
={\bf b}\lrcorner d\boldsymbol{\gamma} +d({\bf b}\lrcorner\boldsymbol{\gamma}). \label{eq:A7}
\end{equation}
Using the results:
\begin{align}
\boldsymbol{\gamma}=&-({\bf u}\times\tilde{\boldsymbol\Gamma}){\bf\cdot}d{\bf x}, 
\quad d\boldsymbol{\gamma}=-\nabla\times({\bf u}\times\tilde{\boldsymbol\Gamma})
{\bf\cdot}d{\bf S}, \nonumber\\
{\bf b}\lrcorner\boldsymbol{\gamma}
=&({\bf b\cdot}\nabla)\lrcorner\left(-{\bf u}
\times\tilde{\boldsymbol\Gamma}{\bf\cdot}d{\bf x}\right)
=-{\bf b\cdot}({\bf u}\times\tilde{\boldsymbol\Gamma}), \nonumber\\
{\bf b}\lrcorner d\boldsymbol{\gamma}=&{\bf b}\times \left[\nabla\times({\bf u}
\times\tilde{\boldsymbol\Gamma})\right] {\bf\cdot}d{\bf x}. \label{eq:A8}
\end{align}
in (\ref{eq:A7}) gives the result:
\begin{equation}
{\cal L}_{\bf b}\left({\bf u}\lrcorner \beta\right)
= {\bf b}\times[\nabla\times ({\bf u}\times\tilde{\boldsymbol\Gamma})]{\bf\cdot}d{\bf x}
-d\left({\bf b\cdot} ({\bf u}\times\tilde{\boldsymbol\Gamma})\right). 
\label{eq:A9}
\end{equation}
Substituting the results (\ref{eq:A9}), (\ref{eq:A7}), (\ref{eq:A5}) in (\ref{eq:A4})
then gives the identity:
\begin{align}
&{\bf b}\times[\nabla\times ({\bf u}\times\tilde{\boldsymbol\Gamma})]{\bf\cdot}d{\bf x}
-d\left({\bf b \cdot} ({\bf u}\times\tilde{\boldsymbol\Gamma})\right)\nonumber\\
&=-[{\bf b},{\bf u}]\times\tilde{\boldsymbol\Gamma}{\bf\cdot}d{\bf x}
+\left\{{\bf u}\times[\nabla\times({\bf b}\times\tilde{\boldsymbol\Gamma})]
-({\bf u}\times{\bf b}) \nabla{\bf\cdot}\tilde{\boldsymbol\Gamma}\right\} {\bf\cdot}
d{\bf x}.  \label{eq:A10}
\end{align}
The result (\ref{eq:A10}), can be written in the form ${\bf Q}{\bf\cdot}d{\bf x}=0$
where
\begin{align}
{\bf Q}=&{\bf b}\times\left[\nabla\times({\bf u}\times\tilde{\boldsymbol\Gamma})\right]
+{\bf u}\times\left[\nabla\times(\tilde{\boldsymbol\Gamma}\times{\bf b})\right]\nonumber\\
&+\tilde{\boldsymbol\Gamma}\times[{\bf u},{\bf b}]
+({\bf u}\times{\bf b})\nabla{\bf\cdot}\tilde{\boldsymbol\Gamma}
+\nabla\left({\bf u\cdot}({\bf b}\times\tilde{\boldsymbol\Gamma})\right)= 0. 
\label{eq:A11}
\end{align}
The result (\ref{eq:A11}) reduces to (\ref{eq:gauge2.22}) for the case 
$\nabla{\bf\cdot}\tilde{\boldsymbol\Gamma}=0$.

\appendix
\section*{Appendix B}
\setcounter{section}{2}
There are different methods that can be used to obtain solutions of the Lie determining equations
(\ref{eq:gauge3.8})-(\ref{eq:gauge3.18a}). In this appendix we use a method that has 
affinities with the steady MHD flows investigated by \cite{Bogoyavlenskij02}, 
\cite{Schief03}, \cite{Golovin10, Golovin11}. These ideas were  used by \cite{Webb05b}
\cite{Webb07} and \cite{WebbMace15} for fluid relabelling symmetries in MHD.
This method allows the function $H$ in (\ref{eq:gauge3.17}) 
to have a  general form involving 
the fluid labels, and the advected invariants.  
Other  solutions of the Lie determining equations are given in Appendix C. 

First consider the solution of (\ref{eq:gauge3.17}), namely:
\begin{equation}
\nabla K+V^\mu\nabla\tilde{\lambda}-V^{\tilde\lambda} \nabla\mu-V^r\nabla S
+{\bf b}\times(\nabla\times V^{\boldsymbol\Gamma})=0. \label{eq:B1}
\end{equation}
Taking the scalar product of (\ref{eq:B1}) with ${\bf b}$ gives the equation:
\begin{equation}
{\bf b\cdot}\nabla K+V^\mu ({\bf b\cdot}\nabla\tilde{\lambda})
- V^{\boldsymbol\lambda} ({\bf b \cdot}\nabla\mu) -V^r ({\bf b\cdot}\nabla S)=0. \label{eq:B2}
\end{equation}
One way in which (\ref{eq:B2}) can be satisfied is if:
\begin{equation}
V^\mu=k_1 {\bf b\cdot}\nabla\mu,\quad  V^{\tilde\lambda}=k_1 {\bf b\cdot}\nabla\tilde{\lambda}, 
\quad {\bf b\cdot}\nabla K=V^r({\bf b\cdot}\nabla S). \label{eq:B3}
\end{equation}
For simplicity, we consider the case:
\begin{equation}
V^r=0\quad\hbox{and}\quad {\bf b\cdot}\nabla K=0. \label{eq:B4}
\end{equation}
Using (\ref{eq:B3})-(\ref{eq:B4}) in (\ref{eq:B1}), (\ref{eq:B1}) reduces to the equation:
\begin{equation}
\nabla K+{\bf b}\times\left[\nabla\times\left(V^{\boldsymbol\Gamma}+k_1 \tilde\lambda\nabla\mu\right)\right]=0. \label{eq:B5}
\end{equation}
Writing 
\begin{equation}
{\bf Q}=\nabla\times\left(V^{\boldsymbol\Gamma}+ k_1\tilde{\lambda}\nabla\mu\right), 
\label{eq:B6}
\end{equation}
equation (\ref{eq:B5}) takes the form:
\begin{equation}
 \bf{Q}\times {\bf b}=\nabla K. \label{eq:B7}
\end{equation}
Equation (\ref{eq:B7}) is reminiscent of the steady MHD form of Faraday's equation
${\bf E}=-{\bf u}\times{\bf B}=-\nabla\psi$ analyzed by 
Schief (2003) and \cite{Bogoyavlenskij02}. 
Note from (\ref{eq:B6}) that
\begin{equation}
\nabla{\bf\cdot}{\bf Q}=\nabla(\rho\hat{\bf Q})=0\quad\hbox{where}\quad \rho \hat{\bf Q}={\bf Q}, \label{eq:B8}
\end{equation}
which resembles the steady mass continuity equation. Similarly,
\begin{equation}
\nabla{\bf\cdot} {\bf B}=\nabla{\bf\cdot}(\rho {\bf b})=0, \label{eq:B9}
\end{equation}
is analogous to the mass continuity equation. 

Taking the curl of (\ref{eq:B7}) gives:
\begin{equation}
\nabla\times\left(\hat{\bf Q}\times{\bf B}\right)=0, \label{eq:B10}
\end{equation}
which is analogous to Faraday's equation $\nabla\times{\bf E}=0$ for steady MHD 
flows (${\bf E}=-{\bf u}\times{\bf B}$). Using (\ref{eq:B8}) in (\ref{eq:B10}) we obtain:
\begin{equation}
\nabla\times\left(\hat{\bf Q}\times{\bf B}\right)=-\rho\left[\hat{\bf Q}, {\bf b}\right]=0, \label{eq:B11}
\end{equation}
where
\begin{equation}
\left[\hat{\bf Q}, {\bf b}\right]=\left[\hat{\bf Q}{\bf\cdot}\nabla{\bf b}
-{\bf b}{\bf\cdot}\nabla \hat{\bf Q}\right] \label{eq:B12}
\end{equation}
is the Lie bracket (commutator) of the vector fields $\hat{\bf Q}$ and ${\bf b}$. Equation
(\ref{eq:B11}) shows that the vector fields (directional derivatives) 
$\hat{\bf Q}$ and ${\bf b}$ define a 2-dimensional, Abelian simple Lie algebra. 
By Frobenius theorem, there exist integral manifolds $\alpha({\bf x})=const.$ and 
$\gamma({\bf x})=const.$ for which $\hat{\bf Q}=\partial{\bf x}/\partial\alpha$ 
and ${\bf b}=\partial {\bf x}/\partial\gamma$ are base vectors which lie in the 
Maxwellian surface  $K=const.$ (i.e. $\alpha=const.$ and $\gamma=const.$  
can be used as generalized coordinates describing the surface). Thus, 
\begin{equation}
\hat{\bf Q}{\bf\cdot}\nabla=\derv{\alpha},\quad {\bf b}{\bf\cdot}\nabla=\derv{\gamma}. 
\label{eq:B13}
\end{equation}
are directional derivatives  $\partial/\partial \alpha$ and $\partial/\partial\gamma$  
in the Maxwell surfaces $K=const.$ (\cite{Schief03}, \cite{Bogoyavlenskij02}, \cite{Webb05b}). 

From \cite{Webb05b}, it follows that 
\begin{equation}
{\bf Q}=\nabla\gamma\times\nabla K,\quad {\bf B}=\nabla K\times\nabla\alpha, \label{eq:B14}
\end{equation}
are solutions of the determining equations for ${\bf Q}$ and ${\bf B}$. From (\ref{eq:B14}): 
\begin{equation}
{\bf Q}\times {\bf b}=\rho \hat{\bf Q}\times{\bf b}
=(\nabla\gamma\times\nabla K)\times{\bf b}
=({\bf b\cdot}\nabla\gamma)\nabla K-({\bf b\cdot}\nabla K)\nabla\gamma=\nabla K, 
\label{eq:B15}
\end{equation}
which verifies  (\ref{eq:B7}) (note that ${\bf b}{\bf\cdot}\nabla\gamma=1$ and 
${\bf b}{\bf\cdot}\nabla K=0$). Using (\ref{eq:B6}) and (\ref{eq:B14}), we obtain:
\begin{equation}
{\bf Q}=\nabla\times\left(V^{\boldsymbol\Gamma}+k_1\tilde{\lambda}\nabla\mu\right)
=\nabla\times(\gamma\nabla K). \label{eq:B16}
\end{equation}
Uncurling (\ref{eq:B16}) we obtain the solution 
\begin{equation}
V^{\boldsymbol\Gamma}=-k_1\tilde{\lambda}\nabla\mu+\gamma\nabla K+k_2\nabla\Lambda, 
\label{eq:B17}
\end{equation}
as a solution for $V^{\boldsymbol\Gamma}$ where $k_2$ is an arbitrary constant, 
and  $\Lambda({\bf x},t)$ is an arbitrary function of ${\bf x}$ and $t$. 
Substitution of the solution  (\ref{eq:B17}) for $V^{\boldsymbol\Gamma}$ in (\ref{eq:gauge3.8}) 
and assuming $d\gamma/dt=0$, we obtain:
\begin{equation}
V^\nu=-k_2\deriv{\Lambda}{t}-k_1\tilde{\lambda}({\bf u \cdot}\nabla\mu)+\gamma ({\bf u}
{\bf\cdot}\nabla K)
=-k_2 \frac{d\Lambda}{dt}+{\bf u}{\bf\cdot} V^{\boldsymbol\Gamma}. \label{eq:B18}
\end{equation}
The solutions (\ref{eq:B17}) and (\ref{eq:B18}) for $V^{\boldsymbol\Gamma}$ and $V^\nu$
are used in Section 4 to obtain MHD conservation laws via Noether's theorem.

\appendix
\section*{Appendix C}
\setcounter{section}{3}

In this appendix, we present solutions of the Lie determining equations 
(\ref{eq:gauge3.8})-(\ref{eq:gauge3.18a}) which in general, are different than 
those presented in Appendix B. 

By assuming that the function $K$ in (\ref{eq:gauge3.18}) has the 
functional form $K=K(\tilde{\lambda},\mu,S)$ (\ref{eq:gauge3.17}) may be reduced to the equation:
\begin{equation}
\nabla\tilde{\lambda} \left(K_{\tilde\lambda}+V^\mu\right)
+\nabla\mu\left(K_\mu-V^{\tilde\lambda}\right)
+\nabla S\left(K_S-V^r\right) +{\bf b}\times\left(\nabla\times V^{\boldsymbol\Gamma}\right)
=0. \label{eq:C1}
\end{equation}
Equation (\ref{eq:C1}) posesses a simple class of solutions of the form:
\begin{align}
V^\mu=&-K_{\tilde\lambda}+d_1 ({\bf b}{\bf\cdot}\nabla\mu) +d_2({\bf b}{\bf\cdot}\nabla S), 
\nonumber\\
V^{\tilde\lambda}=&K_\mu+d_1 ({\bf b\cdot}\nabla{\tilde\lambda}) -d_3 ({\bf b\cdot}\nabla S), 
\nonumber\\
V^r=&K_S+d_2 ({\bf b}{\bf\cdot}\nabla\tilde{\lambda}) +d_3 ({\bf b}{\bf\cdot}\nabla\mu), 
\nonumber\\
V^{\boldsymbol\Gamma}=&-\left[ d_1\tilde{\lambda}\nabla\mu+d_2\tilde{\lambda}\nabla S
+d_3\mu \nabla S\right] +\chi {\bf A}+\nabla\Lambda, \label{eq:C2}
\end{align}
where $d_1$, $d_2$, $d_3$, and $\chi$ are constants and ${\bf B}=\nabla\times{\bf A}$. 

The solution for $V^\nu$ satisfying (\ref{eq:gauge3.8}) has the form:
\begin{equation}
V^\nu={\bf u}{\bf\cdot}\left[ -d_1\tilde{\lambda}\nabla\mu -d_2\tilde{\lambda}\nabla S
-d_3\mu \nabla S\right] +\chi \psi-\Lambda_t. \label{eq:C3}
\end{equation}
This class of solutions are different from the solutions in Appendix B, where the constraints 
$V^r=0$ and ${\bf B}{\bf\cdot}\nabla K=0$ were imposed.  To derive 
the solutions for $V^\mu$, $V^{\tilde\lambda}$, and $V^r$, we first took the 
scalar product of (\ref{eq:C1}) with ${\bf b}$ to determine the effect of 
compatibility conditions parallel to ${\bf b}$. 

In the derivation of (\ref{eq:C1})-(\ref{eq:C3}), there is a critical equation: 
\begin{equation}
\nabla\times V^{\boldsymbol\Gamma}= -\left\{ d_1(\nabla\tilde{\lambda}\times\nabla\mu) 
+d_2(\nabla\tilde{\lambda}\times\nabla S)
+d_3 (\nabla\mu\times\nabla S)\right\} +\chi {\bf B}. \label{eq:C4}
\end{equation}
In ({\ref{eq:C4}),  $d_1$, $d_2$, $d_3$, $\chi$ are not necessarily constants,
in which case it is necessary to uncurl (\ref{eq:C4}).

The solutions (\ref{eq:C1})-(\ref{eq:C3}) may 
give the magnetic helicity, cross helicity  
and  arbitrary potential $\Lambda ({\bf x},t)$ conservation laws by appropriate choice of 
the parameters in Noether's theorem.   

\appendix
\section*{Appendix D}
\setcounter{section}{4}
In this appendix, we prove proposition 5.1. 
We write:
\begin{equation}
F(\rho,S,\mu,{\bf B}; \phi,\beta,\lambda,{\boldsymbol{\Gamma})=\bar F}(\rho,S,{\bf u},{\bf B}), 
\label{eq:gauge5.19d4}
\end{equation}
taking into account the constraints.  
 The transformation of variational derivatives 
may be effected by noting that:
\begin{align}
&\int\biggl(F_\rho\delta\rho +F_\phi\delta\phi+F_S\delta S+F_\beta \delta\beta 
+F_\mu\delta\mu+F_\lambda\delta\lambda +F_{\bf B}{\bf\cdot}\delta {\bf B}
+ F_{\boldsymbol{\Gamma}}{\bf\cdot}\delta\boldsymbol{\Gamma}\biggr) d^3x\nonumber\\
&=\int \left({\bar F}_\rho\delta\rho+{\bar F}_S\delta S+{\bar F}_{\bf u}{\bf\cdot}\delta{\bf u}
+{\bar F}_{\bf B}{\bf\cdot}\delta {\bf B}\right) d^3x. \label{eq:gauge5.19d5}
\end{align}
Using (\ref{eq:gauge2.5}) or (\ref{eq:gauge5.11}) 
to determine $\delta{\bf u}$  
 in (\ref{eq:gauge5.19d5}), integrating by parts and dropping surface terms gives the 
formulae:
\begin{align}
F_\rho=&{\bar F}_\rho+\frac{{\bar F}_{\bf u}}{\rho}{\bf\cdot}
\biggl[r\nabla S
+\tilde{\lambda}\nabla\mu+\tilde{\boldsymbol{\Gamma}}\times{\bf b} 
+\boldsymbol{\Gamma} \frac{\nabla{\bf\cdot}{\bf B}}{\rho}\biggr], \nonumber\\
F_\phi=&-\nabla{\bf\cdot} {\bar F}_{\bf u},
\quad F_S={\bar F}_S+\nabla{\bf\cdot}(r{\bar F}_{\bf u}),
\quad F_\beta=-\frac{{\bar F}_{\bf u}{\bf\cdot}\nabla S}{\rho}, \nonumber\\
F_\mu=&\nabla {\bf\cdot}(\tilde{\lambda} {\bar F}_{\bf u}),
\quad F_\lambda=-\frac{{\bar F}_{\bf u}{\bf\cdot}\nabla\mu}{\rho},\nonumber\\
F_{\boldsymbol{\Gamma}}=&\nabla\times ({\bar F}_{\bf u}\times{\bf b})-{\bar F}_{\bf u}
\frac{\nabla{\bf\cdot}{\bf B}}{\rho}, \nonumber\\
F_{\bf B}=&{\bar F}_{\bf B}
+\nabla\left(\frac{\boldsymbol{\Gamma}{\bf\cdot}{\bar F}_{\bf u}}{\rho}\right)
+\frac{\tilde{\boldsymbol{\Gamma}}\times {\bar F}_{\bf u}}{\rho}. \label{eq:gauge5.19d6}
\end{align}

From (\ref{eq:gauge5.9})-(\ref{eq:gauge5.10}) the variations of $\delta\rho$, $\delta S$ 
$\delta\mu$ and $\delta {\bf B}$ are related to the $F$ and ${\bar F}$ variations by the formulae:
\begin{align}
\delta\rho=&F_\phi=-\nabla{\bf\cdot}{\bar F}_{\bf u}=0,
\quad \delta S=F_\beta=-\frac{{\bar F}_{\bf u}{\bf\cdot}\nabla S}{\rho}=0, \nonumber\\
\delta\mu=&F_{\lambda}=-\frac{{\bar F}_{\bf u}{\bf\cdot}\nabla\mu}{\rho}=0, \nonumber\\
\delta{\bf B}=&F_{\boldsymbol{\Gamma}}=\nabla\times ({\bar F}_{\bf u}\times{\bf b}) 
-{\bar F}_{\bf u}\frac{\nabla{\bf\cdot}{\bf B}}{\rho}=0. \label{eq:gauge5.19d7}
\end{align}
In  (\ref{eq:gauge5.19d4})-(\ref{eq:gauge5.19d7}) we have dropped reference to 
the index $k$ which would apply if there are several Clebsch Lin constraint 
variables $(\mu^k,\lambda^k)$. The basic idea can be illustrated with one 
Lin constraint Clebsch pair. 

Equations (\ref{eq:gauge5.19d7}) apply to  fluid 
relabelling symmetries (e.g \cite{Padhye96a, Padhye96b}). 
Equations (\ref{eq:gauge5.19d7}) may  be written as:
\begin{align}
\delta\rho=&-\nabla{\bf\cdot}\left(\rho \hat{V}^{\bf x}\right),
\quad \delta S=-\hat{V}^{\bf x}{\bf\cdot}\nabla S, \nonumber\\
\delta {\bf B}=&\nabla\times\left(\hat{V}^{\bf x}\times {\bf B}\right)
-\hat{V}^{\bf x} \nabla{\bf\cdot}{\bf B}=0, \label{eq:gauge5.19d8}
\end{align}
where
\begin{equation}
\hat{V}^{\bf x}=\frac{1}{\rho} {\bar F}_{\bf u}\equiv {\bar F}_{\bf M}, \label{eq:gauge5.19d9}
\end{equation}
defines the canonical Lie symmetry operator (vector field): 
\begin{equation}
\hat{V}^{\bf x}=\hat{V}^{x^i}\derv{x^i},\quad 
\hat{V}^{x^i}=V^{x^i}-V^{x_0^s} \deriv{x^i}{x_0^s},  \label{eq:gauge5.19d10}
\end{equation}
associated with the Lagrangian map ${\bf x}={\bf x}({\bf x}_0,t)$ between the Lagrangian 
fluid labels ${\bf x}_0$ and the Eulerian position ${\bf x}$ of the fluid element at time $t$ 
(e.g. \cite{Newcomb62}, \cite{Webb05b}, \cite{Webb07}), where 
$d{\bf x}/dt={\bf u}({\bf x},t)$ is formally integrated to give the solution 
${\bf x}={\bf x}({\bf x}_0,t)$ where ${\bf x}={\bf x}_0$ at time $t=0$. For fluid 
relabeling symmetries $V^{x^i}=0$ in (\ref{eq:gauge5.19d10}) (\cite{Webb07}). 
In (\ref{eq:gauge5.19d9})
${\bar F}_{\bf M}=\delta {\bar F}/\delta {\bf M}$ is the variational derivative of ${\bar F}$ 
with respect to the mass flux ${\bf M}=\rho {\bf u}$ (i.e. $\rho$ and ${\bf M}$ 
are regarded as independent variables rather than $\rho$ and ${\bf u}$). 

Casimirs $C$, are functionals which have zero non-canonical Poisson brackets with all other 
functionals, $F$ defined on the system, i.e. 
\begin{equation}
\{C,F\}=0, \quad \hbox{for all}\quad \hbox{functionals F}
\label{eq:gauge5.19d10a}
\end{equation}
(e.g. \cite{Holm85}, \cite{Hameiri04}).  

Using (\ref{eq:gauge5.19d6})-(\ref{eq:gauge5.19d7}) $F_\rho$, $F_\mu$ and $F_S$ reduce to:
\begin{align}
F_\rho=&{\bar F}_\rho+\frac{{\bar F}_{\bf u}}{\rho} {\bf\cdot}
\left(\tilde{\boldsymbol{\Gamma}}\times{\bf b}+\boldsymbol{\Gamma}
\frac{\nabla{\bf\cdot}{\bf B}}{\rho}\right), \nonumber\\
F_\mu=&{\bar F}_{\bf u}{\bf\cdot}\nabla\tilde{\lambda},\quad F_S
={\bar F}_S+{\bar F}_{\bf u}{\bf\cdot}\nabla r. \label{eq:gauge5.19d11}
\end{align}
The condition (\ref{eq:gauge5.19d8}) that $\delta {\bf B}=0$ may be written in 
the Lie bracket form:
\begin{equation}
\rho[{\bf b},{\bar F}_{\bf M}]\equiv \rho({\bf b}{\bf\cdot}\nabla {\bar F}_{\bf M}
-{\bar F}_{\bf M}{\bf\cdot}\nabla{\bf b})=0. \label{eq:gauge5.19d12}
\end{equation}

To obtain the Casimir determining equation (\ref{eq:gauge5.19d1}), 
note that (\ref{eq:gauge5.16}) and (\ref{eq:gauge5.19d6})-(\ref{eq:gauge5.19d11}) 
together give (\ref{eq:gauge5.16}) in the form:
\begin{align}
&\rho\nabla\left\{
{\bar F}_\rho+\frac{{\bar F}_{\bf u}}{\rho}{\bf\cdot}
\left(\tilde{\boldsymbol{\Gamma}}\times {\bf b}+\boldsymbol{\Gamma}
\frac{\nabla{\bf\cdot}{\bf B}}{\rho}\right)
\right\}
\nonumber\\
&-\left({\bar F}_{\bf u}{\bf\cdot}\nabla\tilde{\lambda}\right)\nabla\mu
-\left({\bar F}_S+{\bar F}_{\bf u}{\bf\cdot}\nabla r\right)\nabla S\nonumber\\
&+{\bf B}\times\left\{
\nabla\times {\bar F}_{\bf B}
+\nabla\times\left(\frac{\tilde{\boldsymbol{\Gamma}}
\times {\bar F}_{\bf u}}{\rho}\right)
\right\}\nonumber\\
&-\left\{{\bar F}_{\bf B}+\nabla\left(\frac{\boldsymbol{\Gamma}{\bf\cdot}{\bar F}_{\bf u}}
{\rho}\right)\right\} \nabla{\bf\cdot}{\bf B}=0. \label{eq:gauge5.19d13}
\end{align}
By using the formulae:
\begin{equation}
\boldsymbol{\omega}=\nabla\times{\bf u}=-\nabla r\times\nabla S
-\nabla\tilde{\lambda}\times\nabla\mu -\nabla\times
\left(\tilde{\boldsymbol{\Gamma}}\times{\bf b}\right)
-\nabla\times\left(\boldsymbol{\Gamma}\frac{\nabla{\bf\cdot B}}{\rho}\right), 
\label{eq:gauge5.19d14}
\end{equation}
and
\begin{equation}
\boldsymbol{\omega}\times{\bar F}_{\bf u}=
-\left({\bar F}_{\bf u}{\bf\cdot}\nabla r\right)
\nabla S
-\left({\bar F}_{\bf u}{\bf\cdot}\nabla\tilde{\lambda}\right)\nabla\mu
+{\bar F}_{\bf u}\times\left[\nabla\times(\tilde{\boldsymbol{\Gamma}}\times{\bf b})
+\boldsymbol{\Gamma} \frac{\nabla{\bf\cdot}{\bf B}}{\rho}\right], \label{eq:gauge5.19d15}
\end{equation}
(\ref{eq:gauge5.19d13}) reduces to:
\begin{equation}
\rho\nabla{\bar F}_\rho-{\bar F}_S\nabla S+{\bf B}\times\left(\nabla\times{\bar F}_{\bf B}\right)
-{\bar F}_{\bf B} \nabla{\bf\cdot}{\bf B} 
+\boldsymbol{\omega}\times {\bar F}_{\bf u}+{\bf R}=0, \label{eq:gauge5.19d16}
\end{equation}
where
\begin{align}
{\bf R}=&\rho\biggl\{\nabla\left[{\bar F}_{\bf M}
{\bf\cdot}\tilde{\boldsymbol{\Gamma}}\times {\bf b} 
+({\bar F}_{\bf M}{\bf\cdot}\boldsymbol{\Gamma})\Theta\right]\nonumber\\
&-{\bar F}_{\bf M}\times\nabla\times\left[\tilde{\boldsymbol{\Gamma}}\times{\bf b}
+\boldsymbol{\Gamma}\Theta\right]
+{\bf b}\times\left[\nabla\times(\tilde{\boldsymbol{\Gamma}}\times {\bar F}_{\bf M})\right]
\nonumber\\
&-\left\{\nabla \left(\boldsymbol{\Gamma}{\bf\cdot}{\bar F}_{\bf M}\right)
+\tilde{\boldsymbol{\Gamma}}\times {\bar F}_{\bf M})\right\}\Theta\biggr\}\nonumber\\
\equiv&
-\rho\biggl\{{\bf b}\times\left[\nabla\times({\bar F}_{\bf M}\times\tilde{\boldsymbol{\Gamma}})
\right]
+{\bar F}_{\bf M}\times[\nabla\times(\tilde{\boldsymbol{\Gamma}}\times {\bf b})]\nonumber\\
&+\nabla[{\bar F}_{\bf M}{\bf\cdot}({\bf b}\times\tilde{\boldsymbol{\Gamma}})]
+\boldsymbol{\Gamma} {\bar F}_{\bf M}{\bf\cdot}\nabla\Theta\biggr\}, \label{eq:gauge5.19d17}
\end{align}
and
\begin{equation}
\Theta=\frac{\nabla{\bf\cdot}{\bf B}}{\rho}. \label{eq:gauge5.19d18}
\end{equation}

By taking the divergence of $\delta {\bf B}$ in (\ref{eq:gauge5.19d8}) we obtain:
\begin{equation}
\nabla{\bf\cdot}\delta {\bf B}=-\nabla{\bf\cdot}[{\bar F}_{\bf M}\nabla{\bf\cdot}{\bf B}]
=-\nabla{\bf\cdot}(\rho {\bar F}_{\bf M}\Theta)=-\rho {\bar F}_{\bf M}{\bf\cdot}\nabla\Theta=0.
\label{eq:gauge5.19d19}
\end{equation} 
Using (\ref{eq:gauge5.19d12}), (\ref{eq:gauge5.19d19}) and (\ref{eq:A11}), 
with ${\bf u}\to {\bar F}_{\bf M}$, we find ${\bf R}=-\rho {\bf Q}=0$ as 
${\bf Q}=0$ for the case ${\bf u}\to {\bar F}_{\bf M}$ in (\ref{eq:A11}).
Equation (\ref{eq:gauge5.19d16}) then reduces to the Casimir determining equation 
(\ref{eq:gauge5.19d1}). This completes the proof of Proposition (\ref{prop5.1}). 

\appendix
\section*{Appendix E}
\setcounter{section}{5}
 In this appendix we discuss the work of \cite{Yahalom13, Yahalom16, Yahalom17a, Yahalom17b} 
on magnetic helicity $H_M$ 
and non-barotropic cross helicity $H_{CNB}$. Yahalom developed a five Clebsch variable variational principle
for MHD (at first sight there appears to be 8 Clebsch variables involved). \cite{Yahalom17a, Yahalom17b} 
uses the action:
\begin{align}
{\cal A}=&\int\biggl\{ \left(\frac{1}{2}\rho u^2 -\rho e(\rho,S)+\frac{B^2}{2\mu_0}\right)\nonumber\\
&+\phi\left[\deriv{\rho}{t}+\nabla{\bf\cdot}(\rho {\bf u})\right] -\rho \alpha \frac{d\chi}{dt}
-\rho\beta \frac{d\eta}{dt} -\rho\sigma \frac{dS}{dt}\nonumber\\
&-\frac{\bf B}{\mu_0}{\bf\cdot}\nabla\chi\times \nabla\eta\biggr\} d^3x\ dt. \label{eq:E1}
\end{align}
The stationary point conditions $\delta{\cal A}/\delta {\bf B}=0$ and $\delta{\cal A}/\delta {\bf u}=0$
gives the Clebsch expansions:
\begin{align}
{\bf B}=&\nabla\chi\times \nabla\eta, \nonumber\\
{\bf u}=&\nabla\phi+\alpha\nabla\chi+\beta\nabla\eta+\sigma\nabla S, \label{eq:E2}
\end{align}
for ${\bf B}$ and ${\bf u}$ (we use $r\equiv -\sigma$ in our formulation).
 It is straightforward to write down the other variational equations by varying 
$\rho$, $S$
and the Clebsch variables in the variational principle (see e.g. \cite{Yahalom17a, Yahalom17b}). 

The magnetic field Clebsch variable expansion (\ref{eq:E2}) is also used by \cite{Sakurai79}. 
The magnetic 
vector potential ${\bf A}$ and ${\bf B}$ have the forms:
\begin{equation}
{\bf A}=\chi\nabla\eta+\nabla \zeta,\quad {\bf B}=\nabla\chi\times\nabla\eta. \label{eq:E3}
\end{equation}
For a non-trivial magnetic field topology there does not exist a global ${\bf A}$ (i.e. $\chi$, $\eta$, $\zeta$ 
are not global single valued functions of ${\bf x}$).  Notice from (\ref{eq:E2}) that the magnetic helicity 
density $h_m={\bf A}{\bf\cdot}{\bf B}$ has the form:
\begin{equation}
h_m={\bf A}{\bf\cdot}{\bf B}=\nabla \zeta{\bf\cdot}\nabla\chi\times \nabla\eta=
\frac{\partial(\zeta,\chi,\eta)}{\partial(x,y,z)}. \label{eq:E4}
\end{equation}
Thus  $h_m\neq 0$ only if $\chi,\eta$ and $\zeta$ are independent functions of ${\bf x}$. 
\cite{Semenov02} argue that the field topology changes due to jumps in $\zeta$ in magnetic fields
with non-trivial topology for generalized versions of the MHD topological 
soliton (see also \cite{Kamchatnov82}).
A similar jump in $\zeta$ occurs in the non-global ${\bf A}$ for the magnetic monopole (\cite{Urbantke03}). 

\cite{Yahalom13,Yahalom17a,Yahalom17b} introduces a further independent magnetic field potential, 
$\mu$ (called the metage)
which represents distance or affine parameter along the magnetic field line formed by the intersection 
of the $\eta=const.$ and $\chi=const.$ Euler potential surfaces. Thus, we obtain:
\begin{equation}
\nabla \zeta=\deriv{\zeta}{\chi}\nabla\chi+\deriv{\zeta}{\eta}\nabla\eta +\deriv{\zeta}{\mu} \nabla\mu. \label{eq:E5}
\end{equation}
Using (\ref{eq:E5}) in (\ref{eq:E4}) gives:
\begin{equation}
h_m={\bf A}{\bf\cdot}{\bf B}=\deriv{\zeta}{\mu}\nabla\mu{\bf\cdot}\nabla\chi\times\nabla\eta=\deriv{\zeta}{\mu} 
\left|\frac{\partial(\chi,\eta,\mu)}{\partial(x,y,z)}\right|. \label{eq:E6}
\end{equation}
The volume integrated magnetic helicity:
\begin{equation}
H_M=\int_V {\bf A}{\bf\cdot}{\bf B}\ d^3x =\int_V \deriv{\zeta}{\mu}\ d\chi\wedge d\eta\wedge d\mu. \label{eq:E7}
\end{equation}
However, the magnetic flux:
\begin{equation}
d\Phi_B={\bf B}{\bf\cdot}d{\bf S}=\left(\nabla\chi\times\nabla\eta\right){\bf\cdot} d{\bf S}=d\chi d\eta. 
\label{eq:E8}
\end{equation}

To prove (\ref{eq:E6}) note that:
\begin{align}
d{\bf S}=&{\bf r}_\chi\times{\bf r}_\eta\ d\chi d\eta\quad\hbox{and}\quad {\bf B}
=\nabla\chi\times\nabla\eta, \nonumber\\
{\bf B}{\bf\cdot}d{\bf S}=&(\nabla\chi\times\nabla\eta) {\bf\cdot}({\bf r}_\chi\times{\bf r}_\eta)\ d\chi d\eta
=d\chi d\eta. \label{eq:E9}
\end{align} 
The last step in (\ref{eq:E9}) follows by setting $(q^1,q^2,q^3)=(\chi,\eta,\mu)$ and noting:
\begin{equation}
\deriv{q^a}{x^k} \deriv{x^k}{q^b}=\delta^a_b\quad \hbox{which implies}\quad {\bf e}^a{\bf\cdot}{\bf e}_b=\delta^a_b, 
\label{eq:E10}
\end{equation}
where ${\bf e}^a=\nabla q^a$ and ${\bf e}_b=\partial {\bf r}/\partial x^b$. 

Using (\ref{eq:E6}) in (\ref{eq:E5}) and integrating over $\mu$ along the field line, we obtain:
\begin{equation}
H_M=\int \left[ \zeta\right]\ d\chi d\eta\equiv \int \left[ \zeta\right]d\Phi_B, \label{eq:E11}
\end{equation}
where $[\zeta]$ is the jump in $\zeta$ between the two ends of the field line (the field lines can be closed 
or open). Equation (\ref{eq:E11}) gives the invariant:
\begin{equation}
[\zeta]=\frac{dH_M}{d\Phi_B}, \label{eq:E12}
\end{equation}
which is the magnetic helicity per unit magnetic flux. (\ref{eq:E12}) shows that for a closed field line, the jump 
in $[\zeta]$ is non-zero for a non-trivial magnetic helicity. \cite{Yahalom13,Yahalom16,Yahalom17a,Yahalom17b} 
refers to  (\ref{eq:E12})  as the MHD `magnetic Aharonov-Bohm effect', in analogy with the Aharonov-Bohm effect
in quantum mechanics. 

\cite{Yahalom13,Yahalom16,Yahalom17a,Yahalom17b} and \cite{Webb14a, Webb14b} developed conservation 
laws for cross helicity and a generalized cross helicity for both barotropic and non-barotropic 
MHD. The cross helicity $H_C$ is is given by:
\begin{equation}
H_C=\int_V {\bf u}{\bf\cdot}{\bf B}\ d^3x. \label{eq:E13}
\end{equation}
The differential form of the cross helicity evolution equation from (\ref{eq:ch6}) is:
\begin{equation}
\derv{t}({\bf u}{\bf\cdot}{\bf B})+ \nabla{\bf\cdot}
\left[({\bf u}{\bf\cdot}{\bf B}){\bf u}+{\bf B}\left(h+\Phi-\frac{1}{2} u^2\right)\right] 
=T({\bf B}{\bf\cdot}\nabla S). \label{eq:E14}
\end{equation}
Integration of (\ref{eq:E14}) over the volume $V$ co-moving with the fluid, and assuming 
${\bf B\cdot n}=0$ on $\partial V$, where ${\bf n}$ is the outward normal to $\partial V$, 
gives the helicity evolution equation:
\begin{equation}
\frac{dH_C}{dt}=\int_V T({\bf B}{\bf\cdot}\nabla S)\ d^3x. \label{eq:E15}
\end{equation}
Thus,  $dH_C/dt=0$ for barotropic flows where $\nabla S=0$. 
For non-barotropic flows, we define the generalized cross helicity as:
\begin{equation}
H_{CNB}=\int_V({\bf u}-\sigma\nabla S){\bf\cdot}{\bf B}\ d^3x, 
\label{eq:E16}
\end{equation}
(in our notation $\sigma=-r$).
Equation (\ref{eq:E14}) then gives:
\begin{equation}
\frac{dH_{CNB}}{dt}=0,\quad \hbox{where}\quad \frac{d\sigma}{dt}=T({\bf x},t). \label{eq:E17}
\end{equation}

Using the Clebsch expansions (\ref{eq:E2}) for ${\bf u}$ and ${\bf B}$, we obtain:
\begin{align}
H_C=&\int {\bf B}{\bf\cdot}\nabla \phi \ d^3x+\int \sigma{\bf B}{\bf\cdot}\nabla S\ d^3x,\nonumber\\
\equiv&\int[\phi] d\Phi_B+\int \sigma \deriv{S}\mu d\mu d\Phi_B. \label{eq:E18}
\end{align}
Also
\begin{equation}
H_{CNB}=H_C-\int\sigma{\bf B}{\bf\cdot}\nabla S\ d^3x=H_C-\int\sigma\deriv{S}{\mu}d\mu d\Phi_B. \label{eq:E19}
\end{equation}
Here $[\phi]$ is the jump in the Clebsch potential across the surface where the multi-valued function $\phi$ 
jumps (for simplicity we assume that there is one such surface, but there could be many such surfaces 
). From (\ref{eq:E18}) and (\ref{eq:E19})
\begin{align}
\frac{dH_C}{d\Phi_B}=[\phi]+\int \sigma\deriv{S}{\mu}d\mu\equiv [\phi]+\oint \sigma dS, \quad 
\frac{dH_{CNB}}{d\Phi_B}= [\phi]. \label{eq:E20}
\end{align}
The net upshot of the analysis is that $dH_{CNB}/d\Phi_B$ is an advected topological invariant (note $d[\phi]/dt=0$ 
follows from the variational equation $\delta{\cal A}/\delta\rho=0$). 
These  results are described in more detail in \cite{Yahalom17a, Yahalom17b}). 

\ack
GMW acknowledges stimulating discussions of MHD 
 and Noether's theorems  with Darryl Holm. 
GMW is supported in part by NASA grant NNX15A165G.
SCA is supported in part by an NSERC grant.

\end{document}